\documentclass[12pt]{article}
\usepackage{pstricks}
\usepackage{color,xcolor}
\usepackage{amssymb,amsmath,bm,bbm}
\usepackage{epsf}
\usepackage{epsfig}
\usepackage{afterpage}
\usepackage{longtable}
\usepackage{cite}
\usepackage{latexsym, mathrsfs}
\usepackage{graphics}
\usepackage{url}
\usepackage{paralist}
\usepackage{bbold}
\usepackage{cleveref}
\usepackage{slashed}

\usepackage{braket}

\newcommand{\spur}[1]{\not\! #1 \,}
\newcommand{\qq}{\quad \quad}

\newcommand{\be}{\begin{equation}}
\newcommand{\ee}{\end{equation}}
\newcommand{\bea}{\begin{eqnarray}}
\newcommand{\eea}{\end{eqnarray}}
\newcommand{\nn}{\nonumber}
\newcommand{\dd}{\displaystyle}

\newcommand{\rvecD}{\overrightarrow D}
\newcommand{\lvecD}{\overleftarrow D}

\begin{document}

\begin{flushright} {BARI-TH/22-733}\end{flushright}

\medskip

\begin{center}
{\Large  Relations among  $B_c \to J/\psi, \eta_c$ form factors }
\\[1.0 cm]
{ {P.~Colangelo$^a$, F.~De~Fazio$^a$,   F.~Loparco$^a$ , N.~Losacco$^{a,b}$, M.~Novoa-Brunet$^{a}$ }
 \\[0.5 cm]}
{\small 
$^a$Istituto Nazionale di Fisica Nucleare, Sezione di Bari,  Via Orabona 4, I-70126 Bari, Italy \\[0.1 cm]
$^b$Dipartimento Interateneo di Fisica ``M. Merlin'', Universit\`a  e Politecnico di Bari, \\ via Orabona 4, 70126 Bari, Italy
}
\end{center}

\vskip 0.8cm

\begin{abstract}
\noindent
We analyze the form factors parametrizing the $B_c \to J/\psi, \eta_c$ matrix elements of the operators in a generalized  low-energy  $b \to c$ semileptonic Hamiltonian.
We consider an expansion in nonrelativistic QCD, classifying the heavy quark spin symmetry breaking terms  and expressing the form factors in terms of universal functions in a selected kinematical range.  Using as an input the   lattice QCD results for the $B_c \to J/\psi$ matrix element of the SM operator,  we obtain information on other  form factors.  The extrapolation to the full kinematical range is also presented.
\end{abstract}

\thispagestyle{empty}


\section{Introduction}
The hadronic uncertainty affecting the description of several weak processes  represents an important limitation for the  Standard Model (SM) predictions. Such a theoretical error needs to  be reduced or (at least) controlled in view of the ongoing and planned  high precision measurements. Moreover, the error assessment is a preliminary step before interpreting the deviations of experimental results from the SM expectations, an important issue considering the  recently observed tensions, the so-called  flavour anomalies. In the time in which 
 the search for phenomena beyond the Standard Model (BSM) through the direct production of new particles at the colliders has not  produced compelling evidence, the search for BSM signals  relies on the precision analysis of virtual effects distorting the SM predictions, as well as on the improved measurements of the fundamental parameters.
 
 In the heavy flavour sector, anomalies have been  detected in charged current  $b \to c \ell \nu$ induced transition of  $B_{d,u}$ mesons, comparing  the  $\ell=e,\,\mu$ with  the $\ell=\tau$ modes   \cite{HFLAV:2019otj,Gambino:2020jvv}.  If such deviations  are due to genuine new physics (NP) phenomena producing violation of lepton flavour universality (LFU), related deviations should be found in  $B_s$, $B_c$ and $b$-baryon decay modes, both exclusive and inclusive \cite{Colangelo:2016ymy,Colangelo:2020vhu}. However, the various  processes are affected by  hadronic uncertainties in different ways, hence it is necessary to analyze each mode separately. 

The present study is devoted to the  exclusive semileptonic $b \to c \ell \nu$ decays of the $B_c$ meson, in particular $B_c \to J/\psi \ell \bar \nu_\ell$ and  $B_c \to \eta_c \ell \bar \nu_\ell$  which are under experimental scrutiny \cite{LHCb:2017vlu}.
For such processes  possible BSM effects can be analyzed in terms of a low-energy  Hamiltonian generalizing the Standard Model one with the inclusion of the full set of dimension-6 operators, as done for other modes 
\cite{Biancofiore:2013ki,Becirevic:2016hea,Alonso:2016oyd,Colangelo:2016ymy,Jung:2018lfu,Colangelo:2018cnj,Murgui:2019czp,Alguero:2020ukk}.  The matrix elements of the various operators in the effective Hamiltonian, parametrized in terms of form factors,  introduce the hadronic uncertainty.  The peculiarity of the processes is that they involve mesons each one comprising two heavy quarks,
strengthening the  interest in $B_c$, a meson with quarkonium structure and  only weak decays. 


The  form factors for the $B_c \to J/\psi$ matrix elements of SM operators have been computed in the full kinematical range of  dilepton invariant mass $q^2$  by  lattice nonrelativistic QCD,  by the HPQCD Collaboration  \cite{Harrison:2020gvo}. Other QCD-based computations employ   QCD sum rules in the low $q^2$ range 
\cite{Colangelo:1992cx,Kiselev:1999sc}.   In the same kinematical range, using standard NRQCD methods the $\langle J/\psi |{\bar c} \Gamma_i b| B_c \rangle$  matrix elements are expressed in the form $\langle J/\psi |{\bar c} \Gamma_i b| B_c \rangle\simeq \psi_{B_c}(0) \psi_{J/\psi}(0) T_i$, in terms of
 meson wave functions at the origin $\psi_{B_c}(0) , \,\psi_{J/\psi}(0)$,  and  of perturbatively calculable hard-scattering kernels $T_i$ describing  spectator interaction corrections and  vertex corrections to factorization    \cite{Qiao:2012vt,Zhu:2017lqu,Shen:2021dat,Tang:2022nqm,Tao:2022yur}.  Perturbative QCD calculations use an analogous principle \cite{Wang:2012lrc,Shen:2014msa,Rui:2016opu,Hu:2019qcn}. Light-Cone sum rules have been applied for all $q^2$ \cite{Leljak:2019eyw}.
These  approaches are affected by their own theoretical uncertainties. As for the determinations based on  quark models, beyond the variation of the  input 
parameters the uncertainty attached to the model can hardly be quantified \cite{Tran:2018kuv,Tang:2020org}.  

 In Ref.\cite{Jenkins:1992nb} it has been observed that the semileptonic $B_c$ form factors  can be expressed in terms of  universal functions in selected kinematical regions, on the basis of the heavy quark spin symmetry for large heavy quark masses. This remark  has prompted  a number of phenomenological analyses of  $B_c$  decays \cite{Colangelo:1999zn,Colangelo:2021dnv,Colangelo:2021myn}.  The relations to the universal functions have been provided at the leading order in the heavy quark mass expansion. Here we  extend the analysis at the next-to-leading order in the expansion, to establish relations among the form factors and a set of universal functions based 
on  the heavy quark spin symmetry and  the power counting rules of nonrelativistic QCD (NRQCD). The relations obtained in such  a systematic expansion 
have two applications. They can be used to test the form factors obtained by different methods, for a quantitative assessment of their theoretical  
uncertainty. Moreover, information can be gained  on form factors that have not been computed yet, which are needed for analyses based on the generalized low-energy Hamiltonian.
For this purpose, the available lattice QCD results  can be employed as  input information. 

We shall use the power counting of NRQCD,  the effective QCD theory relevant for mesons comprising two heavy quarks \cite{Lepage:1992tx,Bodwin:1994jh}. A classification of the NRQCD operators is in \cite{Gunawardana:2017zix}. Using this effective theory and  the  potential-NRQCD  effective theory (pNRQCD) formulated  at a lower scale,   fundamental QCD parameters have been precisely computed, namely the beauty and charm quark mass and the coupling constant $\alpha_s$ \cite{Brambilla:2004jw}. Precise  determinations of  the $B_c$ mass \cite{Brambilla:2001fw,Brambilla:2001qk} and lifetime have also  been obtained  \cite{Aebischer:2021ilm,Aebischer:2021eio}.  Here we focus on semileptonic form factors based on a systematic expansion.

The plan of the paper is as follows. In Sec.\ref{heff} we introduce the $b \to c \ell \nu$ generalized  low-energy Hamiltonian in terms of $D=6$ operators. In Sec.\ref{formalism} we describe the formalism of the NRQCD expansion and in Sec.\ref{sec:expansion} we define the universal functions.  In Sec.\ref{numerics} we present a set of numerical results. In App. \ref{app0} we define the parametrization used for the hadronic matrix elements.
In App. \ref{appA} we provide the form factors in terms of universal functions. In App. \ref{appB} we give the results of the fit of some form factors obtained through the universal functions.

\section{Generalized $b \to c$ semileptonic effective Hamiltonian and $B_c \to  J/\psi, \eta_c$ form factors}\label{heff}
The low-energy Hamiltonian  comprising  the full set of  $D=6$ semileptonic  $b \to c$ operators with left-handed neutrinos can be written in the form:
\bea
H_{\rm eff}^{b \to c \ell \bar \nu}&=& \frac{G_F}{\sqrt{2}} V_{cb} \Big[(1+\epsilon_V^\ell) \left({\bar c} \gamma_\mu  (1-\gamma_5) b \right)\left(  \bar \ell  \gamma^\mu  (1-\gamma_5)   \nu_{\ell } \right)
+  \epsilon_R^\ell \left({\bar c} \gamma_\mu (1+\gamma_5) b \right)\left( \bar \ell  \gamma^\mu  (1-\gamma_5) \nu_{\ell } \right)  \nn \\
&+& \epsilon_S^\ell \, ({\bar c} b) \left( {\bar \ell} (1-\gamma_5) \nu_{\ell } \right)
+ \epsilon_P^\ell \, \left({\bar c} \gamma_5 b\right)  \left({\bar \ell} (1-\gamma_5)\nu_{\ell } \right)  
+ \epsilon_T^\ell \, \left({\bar c}   \sigma_{\mu \nu} (1-\gamma_5) b\right) \,\left( {\bar \ell}   \sigma^{\mu \nu} (1-\gamma_5) \nu_{\ell }\right)     \Big]  . \nn \\ \label{hamil} 
\eea
$G_F$ is the Fermi constant and $V_{cb}$ an element of the Cabibbo-Kobayashi-Maskawa (CKM) mixing matrix.
  ${\cal O}_{SM}=4 (\bar c_L \gamma^\mu b_L) \left( {\bar \ell_L} \gamma_\mu \nu_{\ell L}\right)$ is the SM operator. The Hamiltonian 
  Eq.~\eqref{hamil} also comprises  
 the operator ${\cal O}_{R}=4 (\bar c_R \gamma^\mu b_R) \left( {\bar {\ell_L}} \gamma_\mu \nu_{\ell L}\right)$, the scalar 
  ${\cal O}_S=\left({\bar c} b\right)\left( {\bar \ell} (1-\gamma_5) \nu_\ell \right)$, pseudoscalar
 ${\cal O}_P=\left({\bar c} \gamma_5 b \right)\left( {\bar \ell} (1-\gamma_5)  \nu_\ell \right)$ and tensor
 ${\cal O}_T=\left({\bar c}  \sigma_{\mu \nu} (1-\gamma_5) b \right)\left( {\bar \ell} \sigma^{\mu \nu}  (1-\gamma_5)  \nu_\ell \right)$ operators.
 The Wilson  coefficients $\epsilon^\ell_{V,R,S,P,T}$ are complex and  lepton-flavour dependent,   in general. Eq.~\eqref{hamil}  reduces to the SM   for $\epsilon^\ell_i=0$.
  \footnote{The operator ${\cal O}_R$ is included in the set of dimension 6 operators with left-handed neutrinos. In the Standard Model Effective Field Theory the only 
 dimension-$6$ operator with  right-handed quark current is nonlinear in the Higgs field \cite{Buchmuller:1985jz,Cirigliano:2009wk,Aebischer:2020lsx}. }
The Hamiltonian  Eq.~\eqref{hamil} has been considered in connection with the anomalies in $B \to D^{(*)} \tau \nu_\tau$ vs $B \to D^{(*)} \ell \nu_\ell$ decays, obtaining   information on  the various operators and  bounding the parameter space of the  Wilson coefficients \cite{Biancofiore:2013ki,Becirevic:2016hea,Alonso:2016oyd,Colangelo:2016ymy,Jung:2018lfu,Colangelo:2018cnj,Murgui:2019czp,Alguero:2020ukk}. Exclusive $B$  semileptonic modes induced by the $b \to u$ transition and inclusive $b-$baryon modes have  been studied analogously \cite{Colangelo:2019axi,Colangelo:2020vhu}.

The  $B_c$ and $J/\psi,\,\eta_c$ matrix elements of the operators in  Eq.~\eqref{hamil} require  hadronic form factors, for which
different parametrizations can be used.
The $B_c \to \eta_c$ matrix elements  of the vector   $\bar Q^\prime \gamma_\mu Q$, scalar 
$\bar Q^\prime Q$,  and  tensor $\bar Q^\prime \sigma_{\mu \nu} Q$  and $\bar Q^\prime \sigma_{\mu \nu} \gamma_5 Q$ currents can be  written  in terms of form factors $f_i$ as
\bea
\langle P(p^\prime)| {\bar Q}^\prime  \gamma_\mu Q| {B_c}(p) \rangle &=&   f_+^{B_c \to P}(q^2) \Big(p_\mu+p_\mu^\prime  - \frac{m_{B_c}^2-m_P^2}{q^2} q_\mu\Big)  + \,f_0^{B_c \to P}(q^2)\frac{m_{B_c}^2-m_P^2}{q^2} q_\mu \,\,,  \nn  \\ 
\langle P(p^\prime)| {\bar Q}^\prime Q| {B_c}(p) \rangle &=& f_S^{B_c \to P}(q^2) \,\,, \nn \\
\langle P(p^\prime)| {\bar Q}^\prime  \sigma_{\mu \nu }Q| B_c(p) \rangle &=& -i \frac{2 f_T^{B_c \to P}(q^2)}{m_{B_c}+m_P} \big(p_\mu p_\nu^\prime-p_\nu p^\prime_\mu \big) \,\,, \label{BctoP} \\ 
\langle P(p^\prime)| {\bar Q}^\prime \sigma_{\mu \nu }\gamma_5 Q| { B_c}(p) \rangle &=& - \frac{2 f_T^{B_c \to P}(q^2)}{m_{B_c}+m_P} \epsilon_{\mu \nu \alpha \beta} \, p^\alpha p^{\prime \beta}, \nn 
\eea
where  $P=\eta_c$,   $q=p -p^\prime$  is the momentum transfer to the lepton pair, and the condition $f_+^{B_c \to P}(0)=f_0^{B_c \to P}(0)$ holds. We use $\epsilon^{0123}=+1$
 and  the relation $\sigma_{\mu \nu} \gamma_5=\frac{i}{2}\epsilon_{\mu \nu \alpha \beta}\,\sigma^{\alpha \beta}$.
  $f_S^{B_c \to P}$ is related to $f_0^{B_c \to P}$:
$f_S^{B_c \to P}(q^2)=\displaystyle \frac{m_{B_c}^2-m_P^2}{m_Q-m_{Q^\prime}}f_0^{B_c \to P}(q^2)$ with  quark  masses $m_Q$  and $m_{Q^\prime}$.
The $B_c \to J/\psi$ matrix elements  can be  parametrized as
\bea
\langle V(p^\prime,\epsilon)|{\bar Q^\prime} \gamma_\mu Q| {B_c}(p) \rangle &=& 
- \frac{2 V^{B_c \to V}(q^2)}{ m_{B_c}+m_V} i \epsilon_{\mu \nu \alpha \beta} \epsilon^{*\nu}  p^\alpha p^{\prime \beta}, \nn \\
\langle V(p^\prime,\epsilon)|{\bar Q^\prime} \gamma_\mu\gamma_5 Q| {B_c}(p) \rangle &=&  (m_{B_c}+m_V) \Big( \epsilon^*_\mu -\frac{(\epsilon^* \cdot q) }{q^2} q_\mu \Big) A_1^{B_c \to V}(q^2) \nn\\
&-& \frac{(\epsilon^* \cdot q)} {m_{B_c}+m_V} \Big( (p+p^\prime)_\mu -\frac{m_{B_c}^2-m_V^2 } {q^2} q_\mu \Big) A_2^{B_c \to V}(q^2)  \nn \\
&+& (\epsilon^* \cdot q)\frac{2 m_V}{q^2} q_\mu A_0^{B_c \to V}(q^2),  \nn  \\
\langle V(p^\prime,\epsilon)|{\bar Q^\prime} \gamma_5 Q| {B_c}(p) \rangle &=&-\frac{2 m_V}{m_Q+m_{Q^\prime}} (\epsilon^* \cdot q) A_0^{B_c \to V}(q^2),  \label{BctoV} \\
\langle V(p^\prime,\epsilon)|{\bar Q^\prime} \sigma_{\mu \nu} Q| { B_c}(p) \rangle &=& T_0^{B_c \to V}(q^2) \frac{\epsilon^* \cdot q } {(m_{B_c}+ m_V)^2} \epsilon_{\mu \nu \alpha \beta} p^\alpha p^{\prime \beta}  \nn \\
&+&T_1^{B_c \to V}(q^2) \epsilon_{\mu \nu \alpha \beta} p^\alpha \epsilon^{*\beta}  + T_2^{B_c \to V}(q^2) \epsilon_{\mu \nu \alpha \beta} p^{\prime \alpha} \epsilon^{*\beta}, 
\nn  \\ 
\langle V(p^\prime,\epsilon)|{\bar Q^\prime} \sigma_{\mu \nu}\gamma_5 Q| { B_c}(p) \rangle &=&  i\, T_0^{B_c \to V}(q^2) \frac{\epsilon^* \cdot q} {(m_{B_c}+ m_V)^2} (p_\mu p^\prime_\nu-p_\nu p^\prime_\mu) \nn \\
 &+& i\, T_1^{B_c \to V}(q^2) (p_\mu \epsilon_\nu^*-\epsilon_\mu^* p_\nu)  +i\,T_2^{B_c \to V}(q^2)(p^\prime_\mu \epsilon_\nu^*-\epsilon_\mu^* p^\prime_\nu) , \nn
\eea
with $V=J/\psi$ and $\epsilon$ the $J/\psi$ polarization vector.  The condition  holds:
\be  
A_0^{B_c \to V}(0)= \frac{m_{B_c} + m_V}{2 m_V} A_1^{B_c \to V}(0)-  \frac{m_{B_c} - m_V}{2 m_V}  A_2^{B_c \to V}(0) . \qq \qq
\ee
The  results  for $V$ and $A_{1,2,0}$ computed in \cite{Harrison:2020gvo} will be exploited in our numerical analysis.

In our study  it is  convenient to use a different basis of form factors defined in Appendix \ref{app0}. For  $B_c \to \eta_c$ the relations between the two basis are:
\bea
\frac{m_{B_c}^2 -m_P^2}{q^2}\Big(f_0(q^2)-f_+(q^2)\Big)&=& \frac{1}{ 2 \sqrt{m_{B_c} m_P}} \Big((m_{B_c}+m_P) h_-(w) - (m_{B_c}-m_P)h_+(w) \Big)\,\, \nn \\
f_+(q^2)&=& \frac{1}{ 2 \sqrt{m_{B_c} m_P}} \Big((m_{B_c}+m_P) h_+(w) - (m_{B_c}-m_P)h_-(w)\Big)  \,\, \label{BcP} \\
f_T(q^2)&=&-\frac{m_{B_c}+m_P}{\sqrt{m_{B_c} m_{P}}} \,h_T(w) \,\,\, ,   \nn
\eea
where $\dd v=\frac{p}{m_{B_c} }$, $\dd v^\prime=\frac{p^\prime}{m_{P} }$ and $w=v \cdot v^\prime$, hence  $q^2=m_{B_c}^2+m_P^2-2 m_{B_c} m_P \, w$.
For $B_c \to J/\psi$  the relations are:
 \bea
 V(q^2)&=& {m_{B_c}+m_V \over 2 \sqrt{m_B m_V}} h_V(w) \,\,\,  \nn \\
 A_1(q^2) &=& \sqrt{m_{B_c} m_V}{w+1 \over m_{B_c}+m_V} h_{A_1}(w)  \,\,\,   \nn \\
 A_2(q^2) &=& { m_{B_c}+m_V \over 2 \sqrt{m_{B_c} m_V}} \left(h_{A_3}(w)+{m_V \over m_B} h_{A_2}(w)\right)   \,\,\,    \nn \\
A_0(q^2) &=& { 1 \over 2 \sqrt{m_{B_c} m_V}} \Big( m_{B_c} (w+1) h_{A_1}(w) -(m_{B_c}-m_V w)h_{A_2}(w)-(m_{B_c}w-m_V)h_{A_3}(w) \Big)  \,\,\, \nn  \\
T_0(q^2)&=&-{(m_{B_c}+m_V)^2 \over m_{B_c}\sqrt{m_{B_c} m_V}} \,h_{T_3}(w) \,\,\,    \label{BcV} \\
T_1(q^2)&=&{m_V \over \sqrt{m_{B_c} m_V}} \,\Big( h_{T_1}(w)+ h_{T_2}(w)\Big) \,\,\,  \nn \\
T_2(q^2)&=&{m_{B_c} \over \sqrt{m_{B_c} m_V}} \,\Big( h_{T_1}(w)- h_{T_2}(w)\Big) \,\,\, ,  \nn
\eea
with  $q^2=m_{B_c}^2+m_V^2-2 m_{B_c} m_V w$. 
The  form factors $h_i$ can be related to a set of universal functions  in a kinematical range close to $w=1$.  For  hadrons  comprising a single heavy quark, this has been done in \cite{Falk:1990yz,Falk:1992wt,Leibovich:1997em}. The modifications for the heavy quarkonium  are discussed in the next sections. 

\section{Expansion of the heavy quark field and  the QCD Lagrangian}\label{formalism}
To construct the heavy quark expansion, the  heavy quark QCD field $Q(x)$ with mass $m_Q$  is written factorizing a fast oscillation mass term:
\be
Q(x)=  e^{-i \, m_Q v \cdot x} \psi(x)=e^{-i \, m_Q v \cdot x} \Big( \psi_+(x)+\psi_-(x) \Big) \label{eq:Q}
\ee
with  $\psi_\pm= P_\pm \psi(x)$ and  $\dd P_\pm=\frac{1\pm\slashed{v}}{2}$. $\psi_+$  is the positive energy component of the field (we use the notation adopted in \cite{Aebischer:2021ilm}).  $v$ is identified  with the heavy meson (quarkonium) 4-velocity with $v^2=1$. The  equation of motion allows us to relate $\psi_-$  to $\psi_+$,
\be 
\psi_-(x)=\frac{1}{2 m_Q +i v \cdot D} i \slashed{ D}_\perp \psi_+(x) \label{eq:psim}
\ee 
where $D_{\perp \mu}= D_\mu -(v \cdot D) v_\mu$. 
In the rest frame $v=(1,0,0,0)$ we have $v \cdot D=D_t$ and $D_{\perp \mu}=(0, D_i)$.

Using \eqref{eq:Q} and \eqref{eq:psim}   $Q(x)$  can be expressed in terms of $\psi_+(x)$,
\be
Q(x)= e^{-i \, m_Q v \cdot x} \Big( 1+ \frac{1}{2 m_Q +i v \cdot D} i \slashed{D}_\perp  \Big) \psi_+(x) , \label{eq:psi}
\ee
a nonlocal expression which can be expanded:
\be
Q(x)= e^{-i \, m_Q v \cdot x} \left( 1+\frac{i \slashed{D}_\perp}{2 m_Q} +\frac{(-i v \cdot D)}{2 m_Q}  \frac{i \slashed{D}_\perp}{2 m_Q} + \dots    \right)\psi_+(x) . \label{eq:10}
\ee
The power counting of the various operators is set within NRQCD: $D_t \sim {\tilde v}^2$, $D_\perp \sim {\tilde v}$ and  $\psi_+\sim  \tilde v^{3/2}$, where    $ \tilde v = | \vec{\tilde v}|\ll 1$ is the relative heavy quark 3-velocity in the hadron rest frame \cite{Bodwin:1994jh}. Therefore, 
the second  term in Eq.~\eqref{eq:10} is  ${\cal O}(\tilde v \times \tilde v^{3/2})$,  the third one is  ${\cal O}(\tilde v^3\times \tilde v^{3/2})$.  From now on, the power $\tilde v^{3/2}$ for each quark field will be omitted in the power counting of the operators.

The  QCD Lagrangian expressed in terms of $\psi_+$ 
\be
{\cal L}_{QCD}={\bar \psi}_+(x) \left( i v \cdot D +i \slashed{ D}_\perp  \frac{1}{2 m_Q +i v \cdot D} i \slashed{ D}_\perp \right)\psi_+(x)
\ee
 can  be expanded:
\bea
{\cal L}_{QCD}&=&{\bar \psi}_+(x) \left(i v \cdot D+\frac{(iD_\perp)^2}{2m_Q} +\frac{g}{4 m_Q}\sigma \cdot G_\perp  +\frac{i \slashed{D}_\perp}{2 m_Q} \frac{(-i v \cdot D)}{2 m_Q}  ( i \slashed{D}_\perp)+ \dots \right) \psi_+(x) \nn \\
&=& {\cal L}_0+{\cal L}_1 + \dots \,\,\, . \label{eq:lagexp}
\eea
 In this expression $G_{\perp \mu \nu}$   is  $G_{\perp \mu \nu}=(g_{\mu \alpha}-v_\mu v_\alpha)(g_{\nu \beta}-v_\nu v_\beta)G^{\alpha \beta}$. 
 In the rest frame  
$G_{\perp \mu \nu}=G_{ij}$ for $\mu=i=1,2,3$ and  $\nu=j=1,2,3$, while the other components  vanish.
The power counting of the  chromoelectric  field components $E_i=G_{0i}$ and of the chromomagnetic ones $B_i=\displaystyle\frac{1}{2}\epsilon_{ijk}G^{jk}$ is $\tilde v^3$ and $\tilde v^4$, respectively \cite{Lepage:1992tx}.

The first and second term  in Eq.~\eqref{eq:lagexp} are   ${\cal O}(\tilde v^2)$ and provide the leading order Lagrangian 
\bea
{\cal L}_0&=& {\bar \psi}_+(x) \Big(i v \cdot D+\frac{(iD_\perp)^2}{2m_Q}\Big) \psi_+(x) 
\eea
giving the equation of motion for $\psi_+(x)$
\be
\left(i v \cdot D+\displaystyle\frac{(iD_\perp)^2}{2m_Q} \right) \psi_+(x)=0 \,\,. \label{eom}
\ee
The third and fourth term in Eq.~\eqref{eq:lagexp} are  ${\cal O}(\tilde v^4)$ and give
the NLO  Lagrangian 
\bea
{\cal L}_1&=&{\cal L}_{1,1}+{\cal L}_{1,2} \nn 
\eea
where
\bea
 {\cal L}_{1,1}&=&{\bar \psi}_+(x) \frac{g \sigma \cdot G_\perp}{4 m_Q} \psi_+(x) \nn \\
 {\cal L}_{1,2}&=&\bar \psi_+(x) \frac{i \slashed{D}_\perp}{2 m_Q} \frac{(-i v \cdot D)}{2 m_Q}  ( i \slashed{D}_\perp) \psi_+(x) .
 \eea
 Using the equation of motion  together with $[i\slashed{D}_\perp , i v \cdot D] = i g \gamma^{\mu} v^{\nu} G_{\mu \nu}$ and $\slashed{D}_\perp \slashed{D}_\perp =D_\perp^2 - \frac{1}{2} g \sigma \cdot G_\perp$, ${\cal L}_{1,2}$ can be expressed in the form
 \bea
  {\cal L}_{1,2}&=&-\frac{1}{4m_Q^2}\Bigg({\bar \psi}_+(x) \left( -\frac{(i D_\perp)^4}{2m_Q}\right) \psi_+(x) +{\bar \psi}_+(x)\frac{g}{2}\sigma \cdot G_\perp \left( -\frac{(i D_\perp)^2}{2m_Q}\right) \psi_(x) \nn \\
  &&+ig v^\alpha {\bar \psi}_+(x) iD_\perp^\sigma G_{ \alpha \sigma} \psi_+(x) +g v^\alpha {\bar \psi}_+(x)i D_{\perp \tau} \sigma^{\tau \sigma}G_{ \alpha \sigma}\psi_+(x) \Bigg) \\
  &=&{\cal L}_{1,2}^{(1)}+{\cal L}_{1,2}^{(2)}+{\cal L}_{1,2}^{(3)}+{\cal L}_{1,2}^{(4)} . \nn
  \eea
${\cal L}_{1,2}^{(2)}$ is of higher order in the $\tilde v$  expansion.
${\cal L}_1$ can be arranged in the form
 \be
 {\cal L}_1={\cal L}_1^A+{\cal L}_1^B   \label{L1}\\
 \ee
 with
 \bea
 {\cal L}_1^A&=&{\cal L}_{1,1}+{\cal L}_{1,2}^{(4)} =\frac{1}{4m_Q}{\bar \psi}_+(x)A_{\tau \sigma}\sigma^{\tau \sigma} \psi_+(x) \label{nonlocal1}\\
 {\cal L}_1^B &=&{\cal L}_{1,2}^{(1)}+{\cal L}_{1,2}^{(3)}=\frac{1}{4m_Q^2}{\bar \psi}_+(x)B \psi_+(x) , \label{nonlocal2}
 \eea
 where in \eqref{nonlocal1} we have factorized the leading $1/m_Q$ power.
To deal with the antiquark, the QCD field $Q(x)$ is written as
\be
Q(x)=  e^{i \, m_Q v \cdot x} X(x)=e^{i \, m_Q v \cdot x} \Big( X_+(x)+X_-(x) \Big)=
 e^{i \, m_Q v \cdot x} \Big( 1+ \frac{1}{2 m_Q -i v \cdot D} i \slashed{D}_\perp  \Big) X_-(x) ,
\ee
with $X_-$ containing the negative energy component.
The QCD Lagrangian written in terms of $X_-$
\be
{\cal L}_{QCD}={\bar X}_-(x) \left(- i v \cdot D + i \slashed{ D}_\perp \frac{1}{2 m_Q -i v \cdot D} i \slashed{ D}_\perp \right) X_-(x)
\ee
is expanded as
\be
{\cal L}_{QCD}={\bar X}_-(x) \left(- i v \cdot D +\frac{(iD_\perp)^2}{2m_Q}  + \dots \right) X_-(x) \label{lag-qbar}.
\ee
The above expressions define the effective theory in which to work out  the meson form factors.

\section{Meson form factors in the effective theory}\label{sec:expansion}
To obtain the meson form factors in the effective theory, we expand
the weak current involving two heavy quarks $\bar Q^\prime \Gamma Q$, with $\Gamma$  a generic Dirac matrix:
\bea
{\bar Q^\prime(x)}\Gamma Q(x)&=&{\bar \psi}^\prime_+(x) \Big( 1-\frac{i {\overleftarrow {\slashed D}'}_\perp }{2 m_{Q^\prime}} -\frac{1}{4m_{Q^\prime}^2}(i {\overleftarrow {\slashed D}'}_\perp )(i v^\prime \cdot {\lvecD}^\prime) +\dots   \Big)  \nn \\   &&\Gamma  \Big( 1+\frac{i {\overrightarrow {\slashed D}}_\perp}{2 m_Q} +\frac{1}{4m_Q^2}(-i v \cdot {\rvecD}) i {\overrightarrow {\slashed D}}_\perp  + \dots \Big) \psi_+(x) 
\eea
where $D_{\perp \mu}^\prime= D_\mu -(v^\prime \cdot D) v_\mu^\prime$. 
Keeping terms up to  ${\cal O}(1/m_Q^2)$, the current can be written as
\be
{\bar Q^\prime}(x)\Gamma Q(x)=J_0+\Big(\frac{J_{1,0}}{2 m_Q}+\frac{J_{0,1}}{2 m_{Q^\prime}}\Big)+\Big(-\frac{J_{2,0}}{4 m_Q^2}-\frac{J_{0,2}}{4 m_{Q^\prime}^2}+\frac{J_{1,1}}{4 m_Q m_{Q^\prime}} \Big) , \label{current}
 \ee
 with $J_i$ terms 
\bea
J_0 &=& {\bar \psi}_+' \Gamma \psi_+  \nn \\
J_{1,0} &=& {\bar \psi}_+' \Gamma i {\overrightarrow {\slashed D}}_\perp \psi_+ \nn \\
J_{0,1} &=& {\bar \psi}_+' \left( -i {\overleftarrow {\slashed D}'}_\perp \right) \Gamma \psi_+ \nn \\
J_{2,0} &=& {\bar \psi}_+' \Gamma \left(i v \cdot \rvecD \right) i {\overrightarrow {\slashed D}}_\perp\psi_+ \\
J_{0,2} &=& {\bar \psi}_+'  i {\overleftarrow {\slashed D}'}_\perp \left( i v' \cdot \lvecD \right) \Gamma \psi_+ \nn \\
J_{1,1} &=& {\bar \psi}_+' \left( -i {\overleftarrow {\slashed D}'}_\perp \right)\Gamma \left( i {\overrightarrow {\slashed D}}_\perp \right) \psi_+ . \nn
\eea
Considering the power counting  in $\tilde v$,  Eq.~\eqref{current}  comprises  terms up to ${\cal O}(\tilde v^3)$. The ${\cal O}(1/m_Q^3)$ terms involving three derivatives have not been included in \eqref{current}, even though they can be of the same order in  $\tilde v$ of some terms appearing in the (nonlocal) corrections  discussed in the following:  we assume that they provide numerically suppressed effects.  

We  have neglected the perturbative $\alpha_s$ corrections. Considering such corrections the short distance expansion of the current in \eqref{current} would contain more operators and a set of matching coefficients would appear:
\be
{\bar Q^\prime}(x)\Gamma Q(x)=\sum_i C_i(\mu,w) (J_0)_i+ \sum_j \left[\frac{B_j(\mu,w)}{2m_b}(J_{1,0})_j+\frac{B_j^\prime(\mu,w)}{2m_c}(J_{0,1})_j \right]+...\,. \label{current-as}
\ee
The various structures  in Eq.~\eqref{current} can be identified with the first terms in each of the sums in \eqref{current-as}.
The coefficients $C_i$ and  $B_j^{(\prime)}$ are perturbatively expanded in $\alpha_s$. Only the coefficients of the operators  in \eqref{current} contribute at leading order in $\alpha_s$, the others  start at ${\cal O}(\alpha_s)$. 
At leading order in the inverse HQ mass expansion and at ${\cal O}(\alpha_s)$ one finds for the various  currents:
\bea
{\bar Q^\prime} Q&=&{\bar \psi}_+' \big( 1+ \frac{\alpha_s}{\pi}\, C_S\big) \psi_+ 	+...	 \nn\\
{\bar Q^\prime}\gamma_5 Q&=&{\bar \psi}_+' \big( 1+  \frac{\alpha_s}{\pi}\, C_P\big) \gamma_5\psi_+ +...\nn\\
{\bar Q^\prime}\gamma_\mu Q&=&{\bar \psi}_+'  \big[\big(1+  \frac{\alpha_s}{\pi}\, C_{V_1} \big) \gamma_\mu + \frac{\alpha_s}{\pi}\, C_{V_2}\, v_\mu +  \frac{\alpha_s}{\pi}\, C_{V_3}\, v^\prime_\mu \big] \psi_+ 	+... \label{tab-currents}\\
{\bar Q^\prime}\gamma_\mu \gamma_5Q&=&{\bar \psi}_+'  \big[\big(1+  \frac{\alpha_s}{\pi}\, C_{A_1} \big) \gamma_\mu +  \frac{\alpha_s}{\pi}\, C_{A_2}\, v_\mu +  \frac{\alpha_s}{\pi}\, C_{A_3}\, v^\prime_\mu \big] \gamma_5 \psi_+ 	+...  \nn \\
{\bar Q^\prime} \sigma^{\mu\nu} Q	&=& {\bar \psi}_+'  \big[\big(1+ \frac{\alpha_s}{\pi}\, C_{T_1} \big) \sigma^{\mu\nu} + \frac{\alpha_s}{\pi}\, C_{T_2}\, i(v^\mu\gamma^\nu - v^\nu\gamma^\mu) 
  +  \frac{\alpha_s}{\pi}\, C_{T_3}\, i(v'^\mu\gamma^\nu - v'^\nu\gamma^\mu) \big] \psi_+ 	+...\nn
\eea
The  coefficients $C_i$ have been computed in \cite{Falk:1990de,Falk:1990cz,Neubert:1992tg},  the results for $B_j^{(\prime)}$ for the vector and axial vector currents are in \cite{Neubert:1993iv}.
 In Sec. \ref{numerics} we  comment on the accuracy of  using \eqref{current} instead of \eqref{current-as}.

The $B_c$ and $J/\psi, \eta_c$  matrix elements of the various terms in  the expansion \eqref{current} can be expressed using the trace formalism \cite{Falk:1991nq}.
In this formalism, the  lowest-lying S-wave $\bar b  c$ and $\bar c c$ bound states are described by $4 \times 4$ matrices  \cite{Jenkins:1992nb}
\bea
H^{c{\bar b}}(v)&=&\frac{1+\slashed{v}}{2} \left[B_c^{*\mu} \gamma_\mu-B_c \gamma_5 \right]\frac{1-\slashed{v}}{2} \label{Bc} \\
 H^{c{\bar c}}(v^\prime)&=&\frac{1+\spur{v^\prime}}{2} \left[\Psi^{*\mu} \gamma_\mu-\eta_c \gamma_5 \right]\frac{1-\spur{v^\prime}}{2} \,\,\label{psi} 
\eea
satisfying the relations  ${\slashed v}H(v)=H(v)=-H(v){\slashed v}$ and $H(v^\prime){\slashed v}^\prime=H(v^\prime)=-{\slashed v}^\prime H(v^\prime)$.
 $B_c^{*\mu},\,B_c$ and $\Psi^{*\mu},\,\eta_c$ annihilate vector and  pseudoscalar $\bar b  c$ and $\bar c c$ mesons of velocity $v$ and $v^\prime$, respectively. They are normalized to $\sqrt{m_M}$ with  $M$ one of the mesons in the spin doublet. The trace formalism has been used to write the effective Lagrangians  governing  strong and radiative heavy quarkonium transitions  in the soft-exchange approximation \cite{Casalbuoni:1992fd,Casalbuoni:1992yd,DeFazio:2008xq}.

The $x$-dependence of the matrix element ${\cal M}_0(x)=\langle M^\prime(v^\prime)|{\bar \psi^\prime}_+(x) \Gamma \psi_+(x) |M(v)\rangle$ can be obtained exploiting the dependence in the effective theory \cite{Mannel:1994xh,Mannel:1994xc}:
\be
|M(x)\rangle=e^{-i {\tilde \Lambda}v \cdot x}|M(0)\rangle \label{states}
\ee
where, for a  $Q{\bar Q}^\prime$ meson, 
\be 
\tilde \Lambda=m_H-m_Q-m_{{\bar Q}^\prime} . \label{eq:lambda}
\ee
Eq.~\eqref{eq:lambda}  means that for  the   heavy quarkonium under scrutiny the  heavy quark binding is generated by nonperturbative effects.   $B_c$ and $J/\psi, \eta_c$ are not considered as purely Coulombic states.
The small binding energy scale $\tilde \Lambda$  can display a residual heavy quark mass dependence. 
From \eqref{states} and \eqref{eq:lambda} we  have
\be
{\cal M}_0(x)=e^{-i \phi \cdot x}{\cal M}_0(0) \label{Ax} 
\ee
with $\phi={\tilde \Lambda}v-{\tilde \Lambda}^\prime v^\prime$. For  $B_c \to J/\psi(\eta_c)$,  the binding energies $\tilde \Lambda^{(\prime)}$ are given by
   ${\tilde \Lambda}=m_{B_c}-m_b-m_c$ and ${\tilde \Lambda}^\prime=m_{J/\psi(\eta_c)}-2m_c$. 

Using the trace formalism, we define
\bea
\langle M^\prime(v^\prime)|{\bar \psi^\prime}_+\Gamma (i v \cdot {\overrightarrow D} ) \psi_+|M(v)\rangle &=& -\phi_K(w) {\rm Tr} \left[ {\bar H}^\prime(v^\prime)\Gamma H(v)\right]\label{kin-en} \\
\langle M^\prime(v^\prime)|{\bar \psi^\prime}_+(-i v^\prime \cdot {\overleftarrow D})\Gamma \psi_+ |M(v)\rangle &=&-\phi_K^\prime(w) {\rm Tr} \left[ {\bar H}^\prime(v^\prime)\Gamma H(v)\right]
\label{kin-en-prime}
\eea
with $w=v \cdot v^\prime$. 
The same formalism allows us to parametrize the matrix elements of the various terms in  (\ref{current}).
The matrix element of $J_0$ is written as
\be
\langle M^\prime(v^\prime)|J_0|M(v) \rangle=-\Delta(w) {\rm Tr} \left[ {\bar H}^\prime(v^\prime)\Gamma H(v)\right] \label{deltaLO}
\ee
and involves the form factor $\Delta(w)$ obtained in \cite{Jenkins:1992nb}.
The  $1/m_{Q,Q^\prime}$ terms involve  the functions $\Delta_\alpha$ and ${\bar \Delta_\alpha}$, 
\bea
\langle M^\prime(v^\prime)|{\bar \psi}^\prime_+\Gamma^\alpha iD_\alpha  \psi_+M(v) \rangle&=&-{\rm Tr} \left[\Delta_\alpha(v,v^\prime) {\bar H}^\prime(v^\prime)\Gamma^\alpha H(v)\right] \label{delta}\\
\langle M^\prime(v^\prime)|{\bar \psi}^\prime_+ (- i{\overleftarrow D}_\alpha)\Gamma^\alpha  \psi_+|M(v) \rangle&=&-{\rm Tr} \left[{\bar \Delta}_\alpha(v,v^\prime)
 {\bar H}^\prime(v^\prime)\Gamma^\alpha H(v)\right] ,
\label{deltabar}
\eea
which  are  expressed  in general as
\bea
\Delta_\alpha(v,v^\prime)&=&\Delta_+(w)(v+v^\prime)_\alpha+\Delta_-(w)(v-v^\prime)_\alpha-\Delta_3(w)\gamma_\alpha \label{deltavv} \\
{\bar \Delta}_\alpha(v,v^\prime)&=&{\bar \Delta}_+(w)(v+v^\prime)_\alpha+{\bar \Delta}_-(w)(v^\prime-v)_\alpha-{\bar \Delta}_3(w)\gamma_\alpha  . \label{deltabarvv} 
\eea
Exploiting the relation 
\be
i\partial_\alpha ({\bar \psi}_+^\prime \Gamma \psi_+)={\bar \psi}_+^\prime ( i{\overleftarrow D}_\alpha)\Gamma \psi_+ +{\bar \psi}_+^\prime \Gamma ( i{\overrightarrow D}_\alpha) \psi_+ \label{deriv}
\ee
and using \eqref{Ax},
 the functions in (\ref{deltavv})-(\ref{deltabarvv}) can be connected to  $\Delta$ in (\ref{deltaLO}):
\bea
({\tilde \Lambda}-{\tilde \Lambda}^\prime w)\Delta&=&(\Delta_+-{\bar \Delta}_+)(1+w)+(\Delta_-+{\bar \Delta}_-)(1-w)+(\Delta_3-{\bar \Delta}_3) \label{eq1} \\
({\tilde \Lambda}w-{\tilde \Lambda}^\prime )\Delta&=&(\Delta_+-{\bar \Delta}_+)(1+w)+(\Delta_-+{\bar \Delta}_-)(w-1)+(\Delta_3-{\bar \Delta}_3)\,\, . \label{eq2} 
\eea
 $\Delta_i$ satisfy the equations
\bea
\Delta_+(1+w)+\Delta_-(1-w)+\Delta_3&=&\phi_K \label{eq3} \\
{\bar \Delta}_+(1+w)+{\bar \Delta}_-(1-w)+{\bar \Delta}_3&=&\phi_K^\prime  \label{eq4} \\
\Delta_3&=&{\bar \Delta}_3
\eea
 obtained using Eqs.~\eqref{eom} and \eqref{Ax}, with  solutions 
\bea
\Delta_+(w)&=&-\frac{\Delta_3(w)}{(1+w)}+\frac{\Delta(w)}{2(1+w)}
\left({\tilde \Lambda}w-{\tilde \Lambda}^\prime\right)+\frac{\phi_K(w)+\phi_K^\prime(w)}{2(1+w)} \\
{\bar \Delta}_+(w)&=&-\frac{\Delta_3(w)}{(1+w)}+\frac{\Delta(w)}{2(1+w)}\left({\tilde \Lambda}^\prime w-{\tilde \Lambda}\right)+\frac{\phi_K(w)+\phi_K^\prime(w)}{2(1+w)} \\
\Delta_-(w)&=&\frac{\Delta(w)}{2(w-1)}\left({\tilde \Lambda}w-{\tilde \Lambda}^\prime\right)-\frac{\phi_K(w)-\phi_K^\prime(w)}{2(w-1)} \\
{\bar \Delta}_-(w)&=&\frac{\Delta(w)}{2(w-1)}\left({\tilde \Lambda}^\prime w-{\tilde \Lambda}\right)+\frac{\phi_K(w)-\phi_K^\prime(w)}{2(w-1)} .  \label{eq:deltasys}
\eea
The functions $\Delta_-$ and $\bar \Delta_-$ are finite at $w=1$ if $\phi_K(w)$ and $\phi_K^\prime(w)$ are related by
the condition 
\be
\phi_K(w)-\phi_K^\prime(w)=\left({\tilde \Lambda}-{\tilde \Lambda}^\prime\right)\Delta(w) \,\,.\label{condition}
 \ee
Solving Eqs.~\eqref{eq1}-\eqref{eq4}   after an expansion of the universal functions close to $w=1$ we find that Eq.~\eqref{condition} holds for $w=1$. There is an interesting  analogy to the case of heavy-light mesons, where $\phi_K(w)=\phi_K^\prime(w)$ and $\phi_K(1)$ is proportional to the heavy quark kinetic energy \cite{Falk:1992wt}. If the relation holds  for all values of $w$,  the heavy-light case  is recovered for  ${\tilde \Lambda}={\tilde \Lambda}^\prime$.
The condition \eqref{condition} allows us to obtain:
\bea
\Delta_+(w)&=&-\frac{\Delta_3(w)}{(w+1)}+\Delta(w)\frac{\tilde \Lambda}{2}\frac{(w-1)}{(1+w)}+\frac{\phi_K(w)}{w+1} \label{rismeno} \\
{\bar \Delta}_+(w)&=&-\frac{\Delta_3(w)}{(w+1)}+\Delta(w)\frac{{\tilde \Lambda}^\prime}{2}\frac{(w-1)}{(1+w)}+\frac{\phi_K^\prime(w)}{w+1}  \label{rispiubar} \\
\Delta_-(w)&=&\frac{\Delta(w)}{2}{\tilde \Lambda} \label{rismeno1} \\
{\bar \Delta}_-(w)&=&\frac{\Delta(w)}{2}{\tilde \Lambda}^\prime\,\,.\label{rismenobar1} 
\eea
The matrix elements of the operators with $1/m_{Q,Q^\prime}$  in (\ref{current}) are expressed in terms of $\Delta_i$:
\bea
\frac{1}{2 m_Q}\langle M^\prime(v^\prime)|J_{1,0}|M(v)\rangle&=&-\frac{1}{2 m_Q}\bigg\{-{\rm Tr} \left[ {\bar H}^\prime(v^\prime)\Gamma H(v)\right]\bigg[\frac{\Delta_3(w)}{1+w}+\frac{w}{1+w}(\phi_K(w)-\Delta(w){\tilde \Lambda})\bigg]\nn \\
&+&{\rm Tr} \left[ {\bar H}^\prime(v^\prime)\Gamma  {\slashed v}^\prime H(v)\right]\bigg[-\frac{\Delta_3(w)}{1+w}+\frac{1}{1+w}(\phi_K(w)-\Delta(w){\tilde \Lambda})\bigg]\nn \\
&-& \left[\gamma^\beta  {\bar H}^\prime(v^\prime)\Gamma \gamma_\beta H(v)\right] \Delta_3(w)\bigg\} \label{J11mat}
\eea
\bea
\frac{1}{2 m_{Q^\prime}}\langle M^\prime(v^\prime)|J_{0,1}|M(v)\rangle &=&-\frac{1}{2 {m_{Q^\prime}}}\bigg\{-{\rm Tr} \left[ {\bar H}^\prime(v^\prime) \Gamma  H(v)\right]\bigg[\frac{\Delta_3(w)}{1+w}+\frac{w}{1+w}(\phi_K(w)-\Delta(w){\tilde \Lambda})\bigg]\nn \\
&+&{\rm Tr} \left[ {\bar H}^\prime(v^\prime) {\slashed v} \Gamma H(v)\right]\bigg[-\frac{\Delta_3(w)}{1+w}+\frac{1}{1+w}(\phi_K(w)-\Delta(w){\tilde \Lambda})\bigg]\nn \\
&-& \left[\gamma^\beta  {\bar H}^\prime(v^\prime)\gamma_\beta \Gamma  H(v)\right] \Delta_3(w)\bigg\}  . \label{J12mat}
\eea

To consider the  $1/m_{Q}^2$ terms we define
\be
\langle M^\prime(v^\prime)|{\bar \psi}^\prime_+ (-i {\overleftarrow D}_\alpha)\Gamma^{\alpha \beta} (i{\overrightarrow D}_\beta)  \psi_+| M(v) \rangle=-{\rm Tr} \left[ \psi_{\alpha \beta}(v,v^\prime) {\bar H}^\prime(v^\prime)\Gamma^{\alpha \beta}  H(v)\right] . \label{eqderLR}
\ee
The function $\psi_{\alpha \beta}(v,v^\prime)$ is written in terms of its   symmetric $\psi^S$ and  antisymmetric  $\psi^A$ parts
\be
\psi_{\alpha \beta}=\frac{1}{2}[\psi_{\alpha \beta}^S+\psi_{\alpha \beta}^A] 
\ee
which can be  parametrized as
\bea
\psi_{\alpha \beta}^S&=& \psi_1^S(w)g_{\alpha \beta}+\psi_2^S(w)(v+v^\prime)_\alpha(v+v^\prime)_\beta+\psi_3^S(w)(v-v^\prime)_\alpha(v-v^\prime)_\beta\nn \\
&+&\psi_4^S(w)\big[(v+v^\prime)_\alpha\gamma _\beta+(v+v^\prime)_\beta\gamma _\alpha \big]+\psi_5^S(w)\big[(v-v^\prime)_\alpha\gamma _\beta+(v-v^\prime)_\beta\gamma _\alpha \big]\nn \\ 
&+&\psi_6^S(w)\big[(v+v^\prime)_\alpha(v-v^\prime)_\beta+(v+v^\prime)_\beta(v-v^\prime)_\alpha \big]  \\
\psi_{\alpha \beta}^A&=& \psi_1^A(w)[v_\alpha v^\prime_\beta-v_\beta v^\prime _\alpha]+\psi_2^A(w)
\big[(v-v^\prime)_\alpha\gamma _\beta-(v-v^\prime)_\beta\gamma _\alpha \big]\nn \\
&+&
\psi_3^A(w)i \sigma_{\alpha \beta}+\psi_4^A(w)\big[(v+v^\prime)_\alpha\gamma _\beta-(v+v^\prime)_\beta\gamma _\alpha \big] .
\eea
The hadronic matrix element of $J_{1,1}$ in (\ref{current}) can be expressed in terms of $\psi_{\alpha \beta}$. For $J_{2,0}$ and $J_{0,2}$ integration by parts is also needed.
We obtain:
\bea
\langle M^\prime(v^\prime)|J_{2,0}|M(v) \rangle&=&-({\tilde \Lambda}-w{\tilde \Lambda}^\prime)\Bigg\{\Delta_+ \Big({\rm Tr} \left[ {\bar H}^\prime(v^\prime)\Gamma  H(v)\right]+{\rm Tr} \left[ {\bar H}^\prime(v^\prime)\Gamma {\slashed v}^\prime H(v)\right] \Big) \nn \\
&+&\Delta_- \Big({\rm Tr} \left[ {\bar H}^\prime(v^\prime)\Gamma  H(v)\right]-{\rm Tr} \left[ {\bar H}^\prime(v^\prime)\Gamma {\slashed v}^\prime H(v)\right] \Big) \nn \\
&-&\Delta_3 {\rm Tr} \left( {\bar H}^\prime(v^\prime)\Gamma \gamma^\beta  H(v) \gamma_\beta\right]-\phi_K {\rm Tr} \left[ {\bar H}^\prime(v^\prime)\Gamma H(v)\right)  \Bigg\} \nn \\
 &-&\frac{1}{2} \Big((1+w)\psi_2^S-(1-w)\psi_3^S-2w\psi_6^S+\psi_1^A-(\psi_4^S-\psi_5^S+\psi_2^A-\psi_4^A) \Big)  \nn \\
 & \times&
 \Big(-w{\rm Tr} \left[ {\bar H}^\prime(v^\prime)\Gamma  H(v)\right]+{\rm Tr} \left[ {\bar H}^\prime(v^\prime)\Gamma {\slashed v}^\prime H(v)\right] \Big)  \\
 & - & \frac{1}{2} \Big((1+w)(\psi_4^S+\psi_4^A)+(1-w)(\psi_5^S+\psi_2^A)+\psi_3^A \Big) \nn \\
 & \times&
 \Big({\rm Tr} \left[ {\bar H}^\prime(v^\prime)\Gamma  H(v)\right]+{\rm Tr} \left[ {\bar H}^\prime(v^\prime)\Gamma \gamma^\beta H(v) \gamma_\beta\right] \Big) , \nn
\eea
\bea
\langle M^\prime(v^\prime)|J_{0,2}|M(v)\rangle&=&({\tilde \Lambda}w-{\tilde \Lambda}^\prime)\Bigg\{{\bar \Delta}_+ \Big({\rm Tr} \left[ {\bar H}^\prime(v^\prime)\Gamma  H(v)\right]+{\rm Tr} \left[ {\bar H}^\prime(v^\prime) {\slashed v} \Gamma H(v)\right] \Big) \nn \\
&+&{\bar \Delta}_- \Big({\rm Tr} \left[ {\bar H}^\prime(v^\prime)\Gamma  H(v)\right]-{\rm Tr} \left[ {\bar H}^\prime(v^\prime){\slashed v} \Gamma  H(v)\right] \Big) \nn \\
&-&{\bar \Delta}_3 {\rm Tr} \left[ \gamma_\beta {\bar H}^\prime(v^\prime)\gamma^\beta\Gamma   H(v)\right]-\phi_K^\prime  {\rm Tr} \left[ {\bar H}^\prime(v^\prime)\Gamma H(v)\right]  \Bigg\} \nn \\
 &-& \frac{1}{2} \Big((1+w)\psi_2^S-(1-w)\psi_3^S+2w\psi_6^S+\psi_1^A-\psi_4^S-\psi_5^S-\psi_2^A-\psi_4^A \Big) \nn  \\
 &\times &
 \Big(-w{\rm Tr} \left[ {\bar H}^\prime(v^\prime)\Gamma  H(v)\right]+{\rm Tr} \left[ {\bar H}^\prime(v^\prime){\slashed v}\Gamma  H(v)\right] \Big)  \\
 &-&\frac{1}{2}  \Big((1+w)(\psi_4^S-\psi_4^A)-(1-w)(\psi_5^S-\psi_2^A)+\psi_3^A \Big) \nn \\
 & \times &
 \Big({\rm Tr} \left[ {\bar H}^\prime(v^\prime)\Gamma  H(v)\right]+{\rm Tr} \left[\gamma_\beta {\bar H}^\prime(v^\prime)\gamma^\beta\Gamma  H(v) \right] \Big)  ,\nn
\eea
\bea
\langle M^\prime(v^\prime)|J_{1,1}|M(v)\rangle&=&
-\frac{1}{2} (\psi_1^S-\psi_3^A) \Big(w{\rm Tr} \left[ {\bar H}^\prime(v^\prime)\Gamma  H(v)\right]-{\rm Tr} \left[ {\bar H}^\prime(v^\prime){\slashed v} \Gamma  H(v)\right] \nn \\
&-&{\rm Tr} \left[ {\bar H}^\prime(v^\prime) \Gamma {\slashed v}^\prime  H(v)\right]
+{\rm Tr} \left[ {\bar H}^\prime(v^\prime)\gamma^\beta \Gamma  \gamma_\beta H(v)\right] \Big)\nn \\
&-&\frac{1}{2} (\psi_2^S-\psi_3^S+\psi_1^A) \Big(w^2{\rm Tr} \left[ {\bar H}^\prime(v^\prime)\Gamma  H(v)\right]\nn \\
&-&w \big({\rm Tr} \left[ {\bar H}^\prime(v^\prime){\slashed v} \Gamma  H(v)\right]+{\rm Tr} \left[ {\bar H}^\prime(v^\prime) \Gamma {\slashed v}^\prime  H(v)\right] \big)
+{\rm Tr} \left[ {\bar H}^\prime(v^\prime){\slashed v} \Gamma {\slashed v}^\prime  H(v)\right] \Big) \nn \\
&-&\frac{1}{2} (\psi_4^S+\psi_2^A+\psi_5^S+\psi_4^A) \Big({\rm Tr} \left[ {\bar H}^\prime(v^\prime)({\slashed v}-w) \Gamma   H(v)  \right] \nn \\
&+&{\rm Tr} \left[ {\bar H}^\prime(v^\prime)({\slashed v}-w) \Gamma  \gamma^\beta H(v) \gamma_\beta \right] \Big)  \\
&-&\frac{1}{2} (\psi_4^S+\psi_2^A-\psi_5^S-\psi_4^A) \Big({\rm Tr} \left[ {\bar H}^\prime(v^\prime) \Gamma ({\slashed v}^\prime-w)   H(v)  \right] \nn \\
&+&{\rm Tr} \left[\gamma^\beta {\bar H}^\prime(v^\prime)\gamma^\beta \Gamma({\slashed v}^\prime-w)    H(v) \right] \Big) \nn \\
&-&\frac{1}{2} \psi_3^A \Big({\rm Tr} \left[ {\bar H}^\prime(v^\prime)\Gamma  H(v)\right]+{\rm Tr} \left[\gamma^\beta  {\bar H}^\prime(v^\prime)\gamma^\beta \Gamma  \gamma_\alpha H(v) \gamma_\alpha\right] \nn \\
&+&{\rm Tr} \left[\gamma_\beta {\bar H}^\prime(v^\prime)\gamma^\beta\Gamma  H(v) \right] 
+{\rm Tr} \left[ {\bar H}^\prime(v^\prime)\Gamma \gamma^\beta H(v)\gamma_\beta \right] \Big) . \nn
\eea
%
Using Eqs.~(\ref{eom}), (\ref{kin-en}) and (\ref{kin-en-prime}) two relations follow:
\bea
2m_Q \phi_K&=&{\tilde \Lambda}^\prime ({\tilde \Lambda}\Delta-\phi_K)(w-1) \nn \\
&-&\frac{1}{2} \big(3\psi_1^S+(1-w^2)(\psi_2^S+\psi_3^S-2\psi_6^S)+2(w-1)(\psi_4^S-\psi_5^S) \big)\label{eqpsi1} \\
2m_{Q^\prime} \phi_K^\prime &=&{\tilde \Lambda} ({\tilde \Lambda}^\prime\Delta-\phi_K^\prime)(w-1) \nn \\
&-&\frac{1}{2} \big(3\psi_1^S+(1-w^2)(\psi_2^S+\psi_3^S+2\psi_6^S)+2(w-1)(\psi_4^S+\psi_5^S)\big) . \label{eqpsi2}
\eea

In addition to the corrections obtained by the expansion of the weak currents, we must consider the corrections to the states. 
Using Eqs.~(\ref{nonlocal1}),(\ref{nonlocal2}), they can be written as
\bea
\langle M^\prime(v^\prime)|i \int d^4x \, T[J_0(0) {\cal L}_1 (x)] |M(v) \rangle&=&-\frac{1}{2m_Q^2} \chi_1(w){\rm Tr} \left[ {\bar H}^\prime(v^\prime)\Gamma  H(v)\right]\nn \\
&-&\frac{1}{4m_Q}{\rm Tr} \left[ \chi_{2 \mu \nu}(w) {\bar H}^\prime(v^\prime)\Gamma  P_+ \left(-\frac{i}{2} \right) \sigma^{\mu \nu}H(v)\right] \qq  \label{NL1} \\
&=&{\cal M}_3+{\cal M}_4 \nn
\eea
and, considering the term analogous to ${\cal L}_1$  for the antiquark in Eq.\eqref{lag-qbar},
\bea
\langle M^\prime(v^\prime)|i \int d^4x \, T[J_0(0) {\cal L}_1^\prime(x) ] |M(v) \rangle&=&-\frac{1}{2m_{Q^\prime}^2}{\bar \chi}_1(w){\rm Tr} \left[ {\bar H}^\prime(v^\prime)\Gamma  H(v)\right]\nn \\ 
&-&\frac{1}{4m_{Q^\prime}}{\rm Tr} \left[{\bar \chi}_{2 \mu \nu}(w) {\bar H}^\prime(v^\prime)\left(-\frac{i}{2} \right)\sigma^{\mu \nu}P_+^\prime \Gamma    H(v)\right] \qq \label{NL2} \\
&=&{\overline {\cal M}}_3+{\overline {\cal M}}_4 \,\, . \nn
\eea
We deal with such corrections parametrizing the functions $\chi_i$, $\bar \chi_i$ as
\bea
\chi_{2 \mu \nu}&=&\chi_2^A i \sigma_{\mu \nu}+\chi_2^B (v_\mu \gamma_\nu-v_\nu \gamma_\mu)+\chi_2^C(v^\prime_\mu \gamma_\nu-v^\prime_\nu \gamma_\mu) \label{chi2} \\
{\bar \chi}_{2 \mu \nu}&=&{\bar \chi}_2^A i \sigma_{\mu \nu}+{\bar \chi}_2^B (v_\mu \gamma_\nu-v_\nu \gamma_\mu)+{\bar \chi}_2^C(v^\prime_\mu \gamma_\nu-v^\prime_\nu \gamma_\mu) . \label{chi2bar} 
\eea
Using ${\slashed v} H(v)=H(v)$ and ${\bar H}^\prime(v^\prime){\slashed v}^\prime={\bar H}^\prime(v^\prime)$ one sees that $\chi_2^B$ and ${\bar \chi}_2^C$ do not contribute to the matrix elements. Hence, ${\cal M}_4$ and $\overline {\cal M}_4$ are given by:
\bea
{\cal M}_4&=&-\frac{1}{4m_Q} \chi_2^A(w) \, d_M {\rm Tr} \left[ {\bar H}^\prime(v^\prime)\Gamma  H(v)\right] \nn \\
&-&\frac{1}{4m_Q} \chi_2^C(w) \, \Big[{\rm Tr} \left[ {\bar H}^\prime(v^\prime)\Gamma  H(v)\right]+{\rm Tr} \left[ \gamma^\nu {\bar H}^\prime(v^\prime)\Gamma P_+ {\slashed v}^\prime \gamma_\nu H(v)\right] \Big]  \label{M4}\\
{\overline {\cal M}}_4&=&-\frac{1}{4m_{Q^\prime}} {\bar \chi}_2^A(w) \,  d_M^\prime {\rm Tr} \left[ {\bar H}^\prime(v^\prime)\Gamma  H(v)\right] \nn \\
&+&\frac{1}{4m_{Q^\prime}} {\bar\chi}_2^B(w) \, \Big[{\rm Tr} \left[ {\bar H}^\prime(v^\prime)\Gamma  H(v)\right]+{\rm Tr} \left[ \gamma^\nu{\bar H}^\prime(v^\prime)\gamma_\nu {\slashed v}  P_+^\prime  \Gamma H(v)\right] \Big]  \,\, , \label{M4bar}
\eea
with  $M=P,V$ and $d_P^{(\prime)}=3,\,d_V^{(\prime)}=-1$.

The above relations  can be used to write the various form factors in terms of universal functions. The resulting expressions are collected in Appendix \ref{appA}.

\section{Numerical results}\label{numerics}

\begin{figure}[!tb]
\begin{center}
\includegraphics[width = 0.43\textwidth]{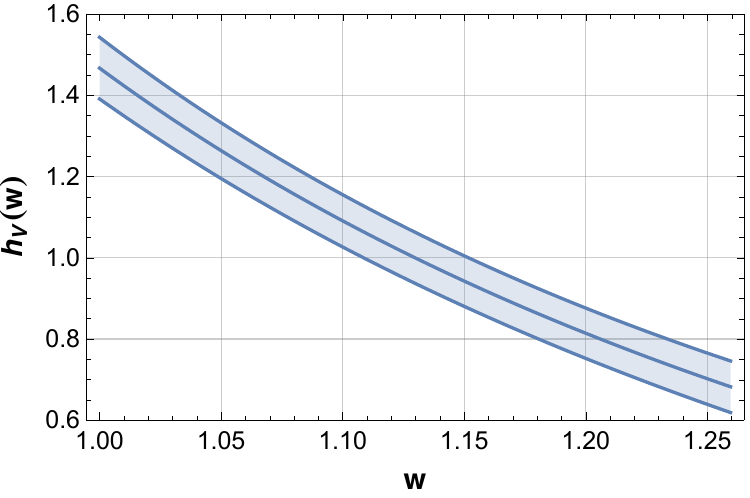}\hskip 0.2cm \includegraphics[width = 0.43\textwidth]{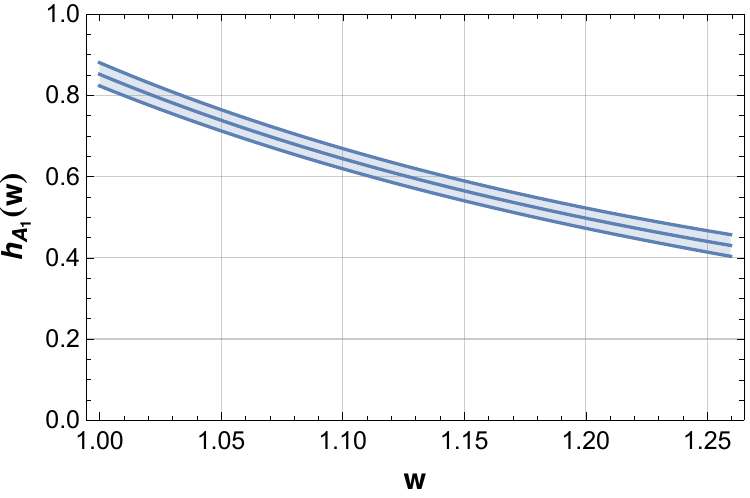}\\
\includegraphics[width = 0.44\textwidth]{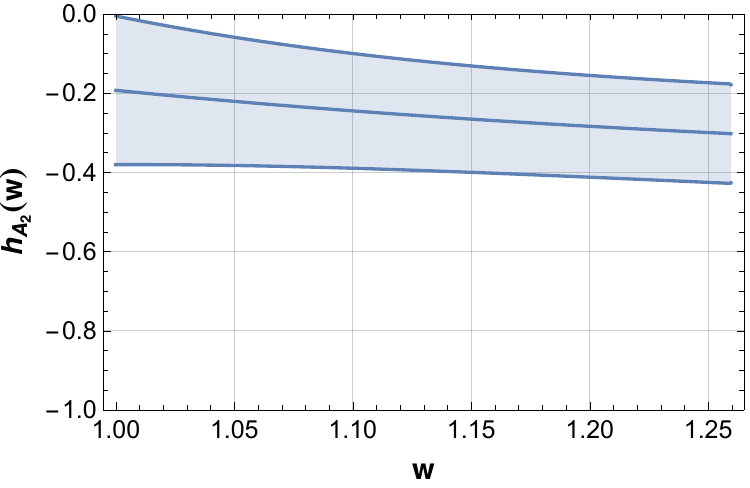}\hskip 0.2cm \includegraphics[width = 0.43\textwidth]{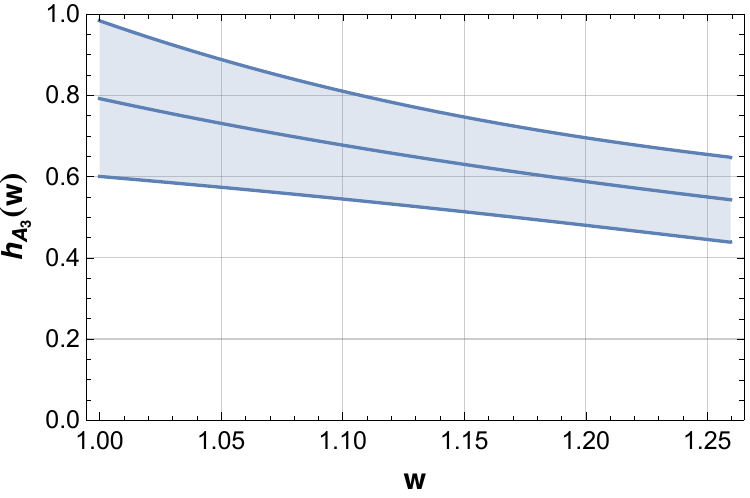}
    \caption{\small $B_c \to J/\psi$ form factors $h_V, \,  h_{A_1},\,h_{A_2}$ and $h_{A_3}$  obtained using the lattice QCD results for  $V,\,A_1,\,A_2,\,A_0$.}\label{hVA}
\end{center}
\end{figure}

\begin{figure}[!tb]
\begin{center}
\includegraphics[width = 0.43\textwidth]{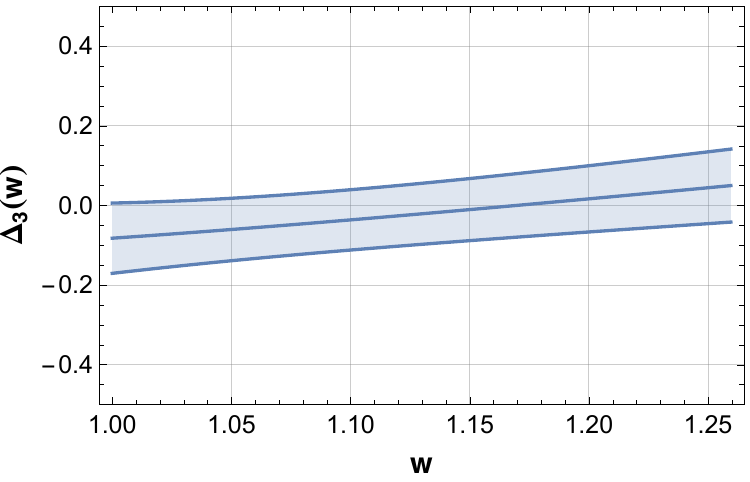}\\
\includegraphics[width = 0.435\textwidth]{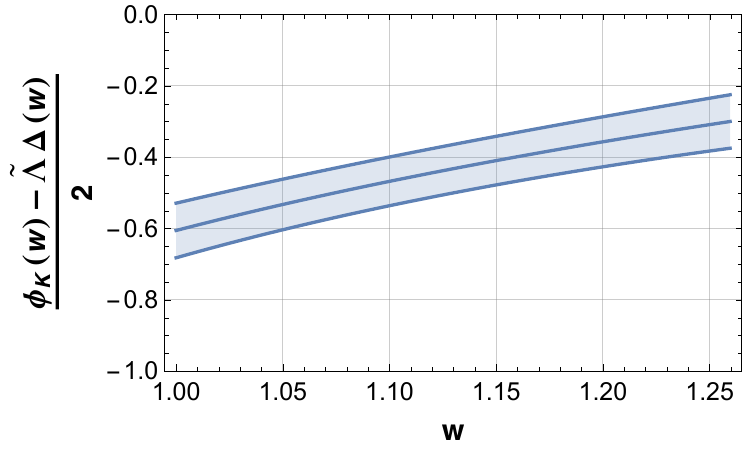}\hskip 0.2cm \includegraphics[width = 0.43\textwidth]{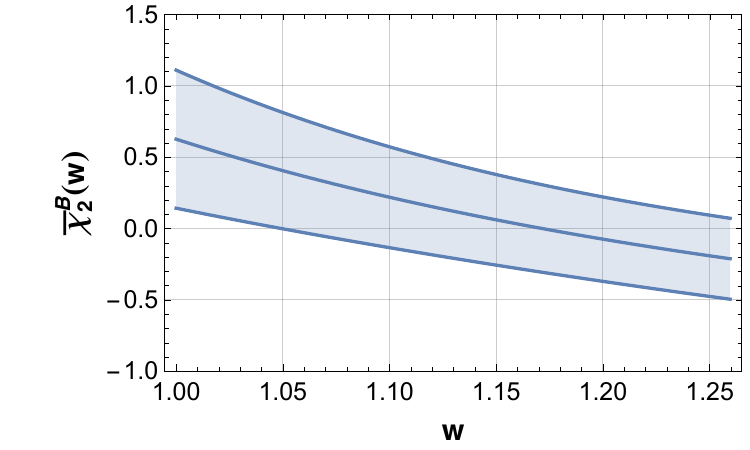}
    \caption{\small  Functions in Eqs.~\eqref{eq:diff}, \eqref{eq:delta3} and \eqref{eq:chiB} extrapolated to the full kinematical range.}\label{univfun}
\end{center}
\end{figure}

\begin{figure}[!tb]
\begin{center}
\includegraphics[width = 0.425\textwidth]{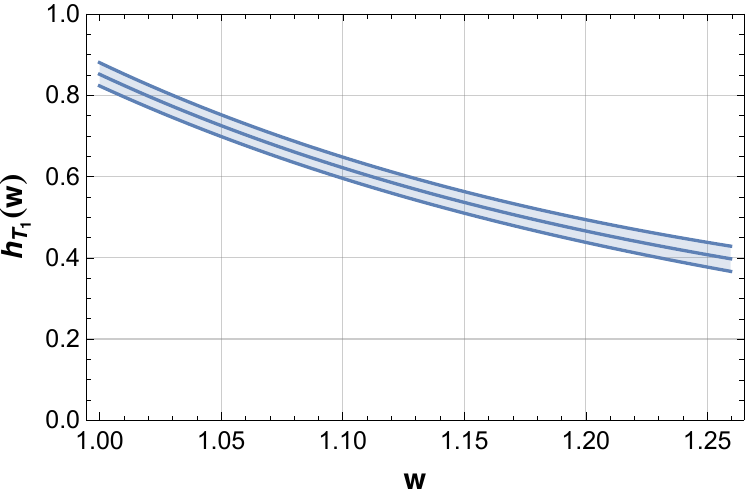}\hskip 0.2cm \includegraphics[width = 0.43\textwidth]{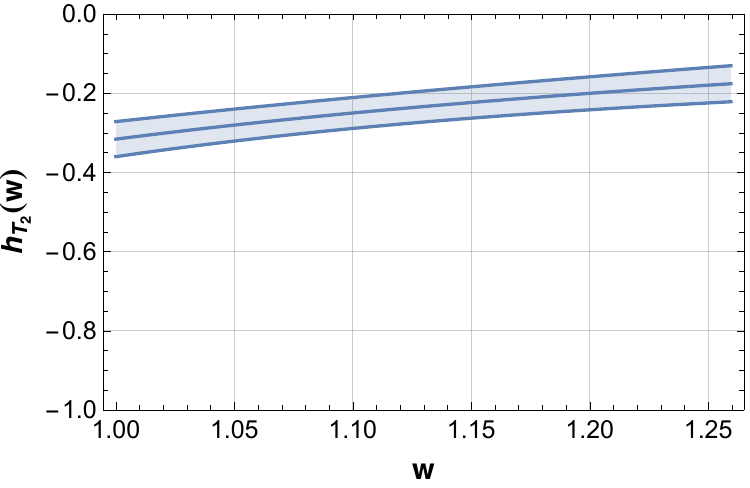}\\
\includegraphics[width = 0.43\textwidth]{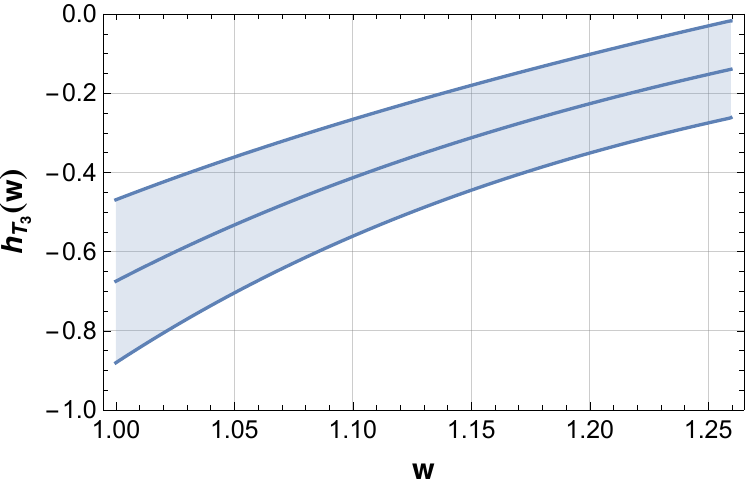}\hskip 0.45cm \includegraphics[width = 0.42\textwidth]{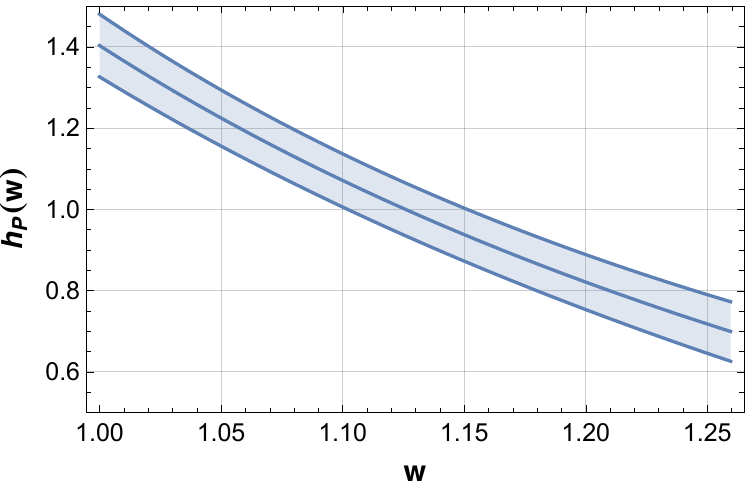}
    \caption{\small Tensor $B_c \to J/\psi$ form factors obtained applying Eqs.~\eqref{eq:hT1}, \eqref{eq:hT2}, \eqref{eq:hT3} and \eqref{eq:hP}   in the full kinematical range   and  using lattice QCD results for  $V$ and $A_{1,2,0}$.   }\label{plotPsiFF}
\end{center}
\end{figure}

\begin{figure}[!tb]
\begin{center}
\includegraphics[width = 0.43\textwidth]{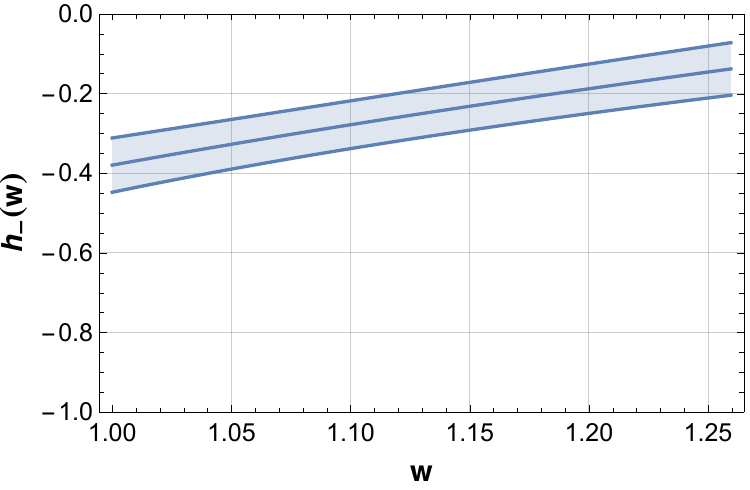}\\
\includegraphics[width = 0.43\textwidth]{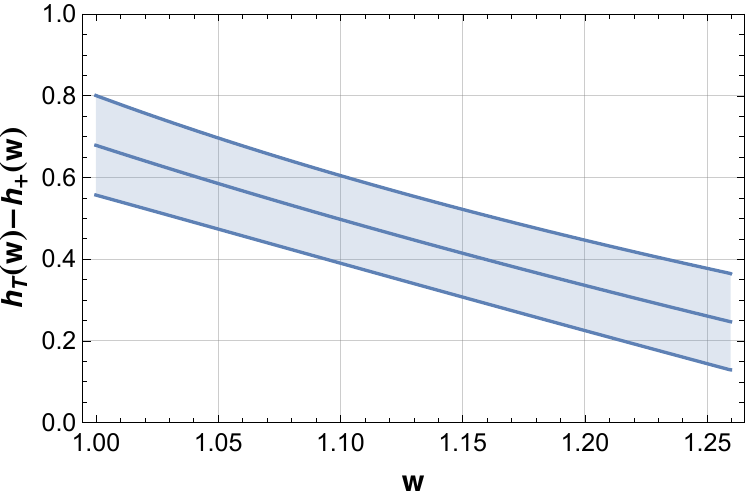}\hskip 0.2cm \includegraphics[width = 0.43\textwidth]{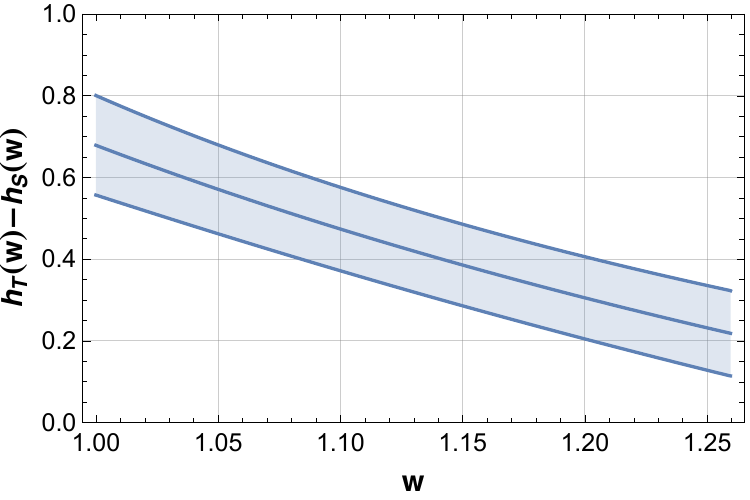}
    \caption{\small  $B_c \to \eta_c$ form factors in Eqs.~\eqref{hmeno}, \eqref{T1menopiu} and \eqref{T1menoS}  extended to the full kinematical range,  obtained using  lattice QCD results for  $V$ and $A_{1,2,0}$.}\label{plotetacFF}
\end{center}
\end{figure}

We present a numerical analysis based on the relations in App.~\ref{appA}.  Since the number of universal functions increases including the various terms, we shall limit the study neglecting  ${\cal O}(1/m^2)$ terms (with $m$ generically $m_b,\,m_c$). This already produces interesting relations. For example, 
the  combinations
\bea
F_1(w)&=&\frac{\phi_K(w)-\Delta(w)\tilde{\Lambda}}{2}+ \Delta_3(w) \label{F1} \\
F_2(w)&=&\frac{\phi_K(w)-\Delta(w)\tilde{\Lambda}}{2}- \Delta_3(w) \label{F2} \\
K(w)&=&3 \chi_2^A(w)+2(w-1)\chi_2^C(w) \label{K} \\
{\bar K}(w)&=&3 {\bar \chi}_2^A(w)-2(w-1){\bar \chi}_2^B(w) \label{Kbar} 
\eea
allow us to simplify the expressions for the form factors and to establish relations among them.
For the $B_c \to \eta_c$ form factors we find:
\bea
h_+(w)&=&\Delta(w)+\frac{1}{4m_b}K(w)+\frac{1}{4m_c}{\bar K}(w)  \nn\\
h_-(w)&=&\left(-\frac{1}{m_b}+\frac{1}{m_c} \right)F_1(w)\nn \\
h_T(w)&=&\Delta(w)-\left(\frac{1}{m_b}+\frac{1}{m_c} \right)F_1(w)
+\frac{1}{4m_b}K(w)+\frac{1}{4m_c} {\bar K}(w) \label{etacFF} \\
h_S(w)&=&\Delta(w)-\left(\frac{1}{m_b}+\frac{1}{m_c} \right)\frac{w-1}{w+1}F_1(w) +\frac{1}{4m_b}K(w)+\frac{1}{4m_c}{\bar K}(w) \,\, . \nn
\eea
For $B_c \to J/\psi$ form factors we obtain:
\bea
h_V(w)&=&\Delta(w)+\frac{1}{4m_b}\left[-4 F_1(w)+ K(w)\right]  -\frac{1}{4m_c}\left[{\bar \chi}_2^A(w)+ 2 (F_1(w)+F_2(w)) \right] \nn \\
h_{A_1}(w)&=&\Delta(w)-\frac{1}{m_b}\frac{w-1}{w+1}F_1(w)-\frac{1}{2m_c}\frac{w-1}{w+1}(F_1(w)+F_2(w)) +\frac{1}{4m_b}K(w)-\frac{1}{4m_c} {\bar \chi}_2^A(w)  
\nn \\
h_{A_2}(w)&=&\frac{1}{m_c}\bigg[{ \frac{1}{2(1+w)} \left(F_1(w)+3 F_2(w) \right)}+\frac{1}{2}{\bar \chi}_2^B(w) \bigg] \nn \\
h_{A_3}(w)&=&\Delta(w)-\frac{1}{m_b}F_1(w) +\frac{1}{4m_b}K(w)-{ \frac{1}{2 m_c(1+w)}\big[wF_1(w)+(w-2)F_2(w)\big]}\nn \\
&-&\frac{1}{4m_c}\big[{\bar \chi}_2^A(w)+2{\bar \chi}_2^B(w)\big]\nn \\
h_{T_1}(w)&=&\Delta(w)+\frac{1}{4m_b}K(w)
-\frac{1}{4m_c}{\bar\chi}_2^A(w) \label{psiFF} \\
h_{T_2}(w)&=&-\frac{1}{m_b}F_1(w)+\frac{1}{2m_c}(F_1(w)+F_2(w))\nn \\
h_{T_3}(w)&=& -\frac{1}{2m_c}\left[ {\bar \chi}_2^B(w) - \frac{1}{1+w} (F_1(w)+3F_2(w))\right]\nn \\
h_P(w)&=&\Delta(w)+\frac{1}{4m_b}\left[-4 F_1(w) + K(w)\right] -\frac{1}{4m_c}\left[{4}F_2(w)+{\bar K}(w) { - 2 {\bar \chi}_2^A (w) }\right] \,\, .  \nn \eea
The lattice results for $V(q^2)$ and $A_{1,2,0}(q^2)$  \cite{Harrison:2020gvo} are translated into  $h_V(w)$ and $h_{A_{1,2,3}}(w)$   in Fig.~\ref{hVA}. 
Keeping the leading term in the heavy quark mass expansion leads to $h_V=h_{A_1}=h_{A_3}=\Delta$ and $h_{A_2}=0$,  relations  badly  violated by the results obtained by lattice QCD, as shown in  Fig.~\ref{hVA}. Therefore, subleading terms must be considered. In this respect,  the same lattice QCD results can be used to predict  other form factors, exploiting Eqs.~\eqref{etacFF} and \eqref{psiFF}.
%
%
A number of  universal functions can be determined. In particular, we find:
\bea
\frac{\phi_K(w)-\Delta(w)\tilde{\Lambda}}{2}&=&\frac{m_c}{2(m_b+3m_c)}(1+w)\bigg(m_b h_{A_1}(w) \nn \\
&+&m_c \left(h_{A_2}(w)+h_{A_3}(w) \right)-(m_b+m_c) h_V(w) \bigg) \label{eq:diff} \\
\Delta_3(w)&=&-\frac{m_c}{2(m_b+3m_c)}(1+w)\bigg(-2m_b h_{A_1}(w)  \nn \\
&+&(m_b+m_c) \left(h_{A_2}(w)+h_{A_3}(w) \right)+(m_b-m_c) h_V(w) \bigg)  \label{eq:delta3}  \\
{\bar \chi}_B^2(w)&=& m_c \left( h_{A_2}(w)-h_{A_3}(w) +h_V(w) \right) .  \label{eq:chiB}
 \eea
 Such functions are displayed in Fig. \ref{univfun}.  In the figure the relations are applied to the full kinematical range, a useful extrapolation for comparing with other calculations.
 This allows us to derive  other form factors  near the zero recoil point. For $B_c \to J/\psi$  the relations hold:
\bea
h_{T_1}(w)&=&\frac{1}{2} \Big((1+w)h_{A_1}(w)-(w-1)h_V(w) \Big) \label{eq:hT1}\\
h_{T_2}(w)&=&\frac{1+w}{2 (m_b + 3 m_c)}\Big((m_b - 3 m_c) h_{A_1}(w) + 2 m_c (h_{A_2}(w) + h_{A_3}(w)) \nn \\
&-& (m_b - m_c) h_V(w)\Big) \label{eq:hT2} \\
h_{T_3}(w)&=& h_{A_3}(w)  -  h_V(w)  \label{eq:hT3}\\
h_P(w)&=&\frac{1}{m_b + 3 m_c}\Big((1+w)\left(m_b h_{A_1}(w)+2m_c h_V(w)\right)\nn \\
&+&\left(-m_b+(w-2)m_c\right)h_{A_2}(w)-\left(w \, m_b+(2w-1)m_c\right)h_{A_3}(w)\Big) \,\, . \label{eq:hP}
\eea
These functions are depicted in Fig.~\ref{plotPsiFF}  extrapolated to the full kinematical range.
For $B_c \to \eta_c$,    $h_-(w)$  and two form factor differences can be derived:
\bea
h_-(w)&=&\frac{m_b - m_c}{2 (m_b + 3 m_c)} (1 + w) \Big(3 h_{A_1}(w) - h_{A_2}(w) - h_{A_3}(w)  - 2 h_V(w)\Big) \label{hmeno} \\
h_T(w)-h_+(w) &=&-\frac{m_b + m_c}{2 (m_b + 3 m_c)} (1 + w) \Big(3 h_{A_1}(w) - h_{A_2}(w) - h_{A_3}(w)  - 2 h_V(w)\Big) \label{T1menopiu} \\
h_T(w)-h_S(w) &=&-\frac{m_b + m_c}{ (m_b + 3 m_c)} \Big(3 h_{A_1}(w) - h_{A_2}(w) - h_{A_3}(w)  - 2 h_V(w)\Big) . \label{T1menoS}
\eea
Their extrapolations are displayed in Fig.~\ref{plotetacFF}. The value of the universal functions at $w=1$ is not predicted, however  from Eqs.~\eqref{eq:hT1} and \eqref{etacFF} the relations $h_{T_1}(w=1)=h_{A_1}(w=1)$ and $h_{S}(w=1)=h_{+}(w=1)$ are obtained, respectively.
%
%
%
It is interesting to observe that if we consider the limit $m_b \to \infty$ keeping  the $1/m_c$ terms, some results remain unaffected. This is the case of Eqs.~\eqref{eq:hT1} and \eqref{eq:hT3}, and of the relations $h_{T_1}(w=1)=h_{A_1}(w=1)$  and $h_S(w=1)=h_+(w=1)$. In the $\eta_c$ case, the relations \eqref{hmeno}-\eqref{T1menoS} are  rescaled  replacing $\displaystyle\frac{m_b\pm m_c}{m_b+3m_c} \to 1$.

The relations among the various form factors are modified using the expansion \eqref{tab-currents}.
In particular,  the right-hand sides of Eqs.~\eqref{eq:hT1}-\eqref{T1menoS}  acquire the extra terms:
\bea
\delta h_{T_1}(w)&=&-\frac{1}{2}\frac{\alpha_s}{\pi}\big[2C_{T_1}+(w-1)\big(C_{T_2}-C_{T_3}+C_{V_1}\big)-(w+1)C_{A_1}\big]\,\Delta(w) \nn \\
\delta h_{T_2}(w)&=&-\frac{1+w}{2}\frac{\alpha_s}{\pi}\big[\big(C_{T_2}+C_{T_3}\big)
+\frac{m_b-m_c}{m_b+3m_c}\big(C_{V_1}-C_{A_1} \big)-\frac{2m_c}{(m_b+3m_c)}\big(C_{A_2}+C_{A_3} \big)\big]\,\Delta(w) \nn \\
\delta h_{T_3}(w)&=&-\frac{\alpha_s}{\pi}\big[C_{T_2}+C_{V_1}-C_{A_1}-C_{A_3} \big] \,\Delta(w) \nn \\
\delta h_P(w)&=&-\frac{\alpha_s}{\pi}\big[C_P-\frac{2m_c}{(m_b+3m_c)}(1+w)C_{V_1}-\frac{m_b-(2w-1)m_c}{(m_b+3m_c)}C_{A_1}
\nn \\
&&
+\frac{m_b-(w-2)m_c}{m_b+3m_c}C_{A_2}+\frac{w \,m_b+(2w-1)m_c}{(m_b+3m_c)}C_{A_3}
\big]\,\Delta(w) \nn \\
\delta h_-(w)&=&-\frac{1+w}{2}\frac{\alpha_s}{\pi}\big[\frac{m_b-m_c}{(m_b+3m_c)}\big(2C_{V_1}-2C_{A_1}+C_{A_2}+C_{A_3}\big) 
+C_{V_2}-C_{V_3}\big]\,\Delta(w) \nn \\
\delta h_{T+}(w)&=&-\frac{1}{2}\frac{\alpha_s}{\pi}\big[2C_{T_1}-2C_{T_2}+2C_{T_3}
-\frac{2[m_b(w+2)+m_c(w+4)]}{m_b+3m_c}C_{V_1} \nn \\ &&
-(1+w)\big(C_{V_2}+C_{V_3} \big) 
+\frac{m_b+m_c}{m_b+3m_c}(1+w)\big(2C_{A_1}-C_{A_2}-C_{A_3}\big)\big]\,\Delta(w) \nn \\
\delta h_{TS}(w)&=&-\frac{\alpha_s}{\pi}\big[C_{T_1}-C_{T_2}+C_{T_3}-C_S+\frac{m_b+m_c}{m_b+3m_c}\big(-2C_{V_1}+2C_{A_1}-C_{A_2}-C_{A_3}\big)\big]\,\Delta(w) \nn \,\,.
\eea
To assess  the size of the new contributions we choose  $w=1$ and $\Delta(1) \simeq1$  \cite{Colangelo:1999zn}. Setting $\mu=\sqrt{m_c\,m_b}$ and  $\alpha_s(\mu)= 0.27$, we find
$\delta h_{T_1}(1) \simeq -0.065$, $\delta h_{T_2}(1) \simeq -0.02$, $\delta h_{T_3}(1)\sim{\cal O}(10^{-4})$, $\delta h_P(1) \simeq 0.025$, $\delta h_-(1) \simeq -0.04$, $\delta h_{T+}(1) \simeq -0.01$, $\delta h_{TS}(1) \simeq -0.09$. Such results compared to the values of the form factors at $w=1$ in Figs. \ref{plotPsiFF}, \ref{plotetacFF}  show that the impact of the matching coefficients is small.

The difference of our approach with  other calculations based on NRQCD must be noticed.  We have used the heavy quark expansion and the NRQCD power counting to relate the various form factors close to the zero recoil point  and exploited lattice QCD to determine, e.g., the tensor form factors. A systematic control of the error can be achieved by the  expansion. Then the results are extrapolated to small  $q^2$. Different  analyses  are carried out at $q^2\simeq 0$, 
where a perturbative approach to the form factor calculation can be attempted,  and extrapolated to higher values of $q^2$ after a normalization to the lattice QCD results  \cite{Tao:2022yur}.
 Numerically, the results match for the form factors $T_1$ and $T_2$ in the basis \eqref{BctoV}, while $T_0$  is affected by a larger uncertainty.
 
 A final remark  is in order about the extrapolation from the kinematical range close to the zero-recoil point $w \sim 1$ to the full kinematical range. There are methods to constrain the extrapolation, starting from the form factors evaluated in few points and using  unitarity constraints and the dispersion matrix \cite{Bourrely:1980gp,Lellouch:1995yv,Boyd:1995sq,
Caprini:1995wq,Boyd:1997kz,Caprini:1997mu,DiCarlo:2021dzg,Martinelli:2021frl}.  The application of such methods to the form factors discussed here is deferred to a dedicated analysis.  Their availability  strengthens the significance of our study,   allowing us to foresee  the control of the hadronic uncertainties not only near the zero-recoil point  but in the full kinematical range.

\section{Conclusions}
Using the heavy quark expansion, the heavy quark spin symmetry and NRQCD power counting  we have  expressed  the form factors parametrizing the matrix elements $\langle J/\psi(\eta_c)|{\bar c} \Gamma_i b|B_c \rangle$ in terms of  universal functions near the zero-recoil point. This can be done for the various operators  in the generalized low-energy Hamiltonian Eq.~(\ref{hamil}), establishing relations among  form factors in a kinematical range around the zero-recoil point.
Lattice QCD results for the matrix element of the Standard Model  operator between $B_c$ and $J/\psi$ allow us to predict the pseudoscalar and tensor form factors. $B_c \to\eta_c$ form factors are also related to the previous ones, obtaining  $h_-$ and the differences between the remaining form factors. We have also presented the results of the extrapolation to the  full kinematical range. The relations worked out in our study can be checked if further information from lattice QCD is available. The effort is to efficiently control the hadronic uncertainties affecting the predictions for semileptonic  $B_c \to J/\psi,\,\eta_c$ decays  in the Standard Model and beyond.

 \vspace*{1.cm}
\noindent {\bf Acknowledgements.}
This study has been  carried out within the INFN project (Iniziativa Specifica) QFT-HEP.

\newpage
\appendix
\numberwithin{equation}{section}
\section{Form factors basis $h_i$ for $B_c \to \eta_c,J/\psi$ }\label{app0}
The  basis of form factors $h_i$ is defined below. \\

\noindent $B_c \to \eta_c$:
\bea
\langle P(v^\prime) | \bar{Q^\prime}  \gamma_\mu  Q | B_c(v) \rangle &=& \sqrt{m_P \, m_{B_c}} \, \big[ h_+(w) \, (v + v')_\mu + h_-(w) \, (v - v')_\mu \big]
\nn \\
\langle P(v^\prime) | \bar{Q^\prime} Q | B_c(v) \rangle &=& \sqrt{m_P \, m_{B_c}} \, h_S(w) (1 + w)  \\
\langle P(v^\prime) | \bar{Q^\prime}  \sigma_{\mu\nu}  Q | B_c(v) \rangle &=&- i \, \sqrt{m_P \, m_{B_c}} \,  h_T(w) \, (v_\mu \, v'_\nu - v_\nu \, v'_\mu)
\nn
\eea
with $P=\eta_c$, $\dd v=\frac{p}{m_{B_c} }$, $\dd v^\prime=\frac{p^\prime}{m_{P} }$ and $w=v \cdot v^\prime$. \\

\noindent $B_c \to J/\psi$:
\bea
\langle V(v^\prime, \epsilon) | \bar{Q^\prime}  \gamma_\mu  Q | B_c(v) \rangle &=&  i \, \sqrt{m_V\, m_{B_c}} \,  h_V(w) \, \epsilon_{\mu\nu\alpha\beta} \, \epsilon^{ *\nu} \, v^{\prime\alpha} \, v^\beta
\nn \\
\langle 
V(v^\prime, \epsilon) | \bar{Q^\prime}  \gamma_\mu  \gamma_5 \, Q | B_c(v) \rangle  &= & \sqrt{m_V \, m_{B_c}} \,  \big[ h_{A_1}(w)  \, (1 + w) \, \epsilon^*_\mu - h_{A_2}(w) \, (\epsilon^* \cdot v) \, v_\mu - h_{A_3}(w) \, (\epsilon^* \cdot v) \, v^\prime_\mu \big]
\nn \\
\langle V(v^\prime, \epsilon) | \bar{Q^\prime}  \gamma_5  Q | B_c(v)  \rangle & =& -\sqrt{m_V \, m_{B_c}}\,  h_P(w) \, (\epsilon^* \cdot v)
 \\
\langle V(v^\prime, \epsilon) | \bar{Q^\prime}  \sigma_{\mu\nu}  Q | B_c(v)  \rangle & =& -\sqrt{m_V\, m_{B_c}}\,\epsilon^{\mu \nu \alpha \beta} \big[ h_{T_1}(w) \epsilon^*_\alpha (v+v^\prime)_\beta+h_{T_2}(w)\epsilon^*_\alpha (v-v^\prime)_\beta\nn \\
&+&h_{T_3}(w)(\epsilon^* \cdot v) v_\alpha v^\prime_\beta \big]
\nn
\eea
with  $V=J/\psi$, $\epsilon$ the $J/\psi$ polarization vector  and $\dd v^\prime=\frac{p^\prime}{m_{V} }$.

\section{Form factors  in terms of universal functions}\label{appA}
The expressions of the form factors $h_i$ in terms of universal functions are given in the following.
\subsection{$B_c \to \eta_c$}
\bea
h_+(w)&=&\Delta(w)+\frac{1}{4m_b}\left(3 \chi_2^A(w)+2(w-1)\chi_2^C(w) \right)+\frac{1}{4m_c}\left(3 {\bar \chi}_2^A(w)-2(w-1){\bar \chi}_2^B(w) \right)  \nn\\
&+&\frac{1}{2m_b^2}\chi_1(w)+\frac{1}{2m_c^2}{\bar \chi}_1(w)  \nn \\
&+&\frac{1}{8m_b m_c}\bigg(-(2+w)\psi_1^S(w)-(w^2-1)\left(\psi_2^S(w)-\psi_3^S(w)+\psi_1^A(w)\right) \label{hpiueta} \\ 
&+&6(w-1)\left(\psi_4^S(w)+\psi_2^A(w) \right)+(w-7)\psi_3^A(w) \bigg) \nn
\eea
\bea
h_-(w)&=&\left(-\frac{1}{m_b}+\frac{1}{m_c} \right)\bigg(\frac{\phi_K(w)-\Delta(w){\tilde \Lambda}}{2}+\Delta_3(w) \bigg)\nn \\
&+&\frac{1}{2m_b^2}\Bigg[({\tilde \Lambda}-w{\tilde \Lambda}^\prime)\bigg(\frac{\phi_K(w)-\Delta(w){\tilde \Lambda}}{2}+\Delta_3(w)\bigg)+\frac{1}{4}(1+w)^2\psi_2^S(w)+\frac{1}{4}(w^2-1)\psi_3^S(w)\nn \\
&+&(1+w)\left(-\psi_4^S(w)-\frac{w}{2}\psi_6^S(w)+\frac{1}{4}\psi_1^A(w)-\frac{1}{2}\psi_4^A(w)\right)\nn \\
&+&\frac{2w-1}{2}\psi_5^S(w)+\frac{w-2}{2}\psi_2^A(w)-\frac{3}{4}\psi_3^A(w)\Bigg] \nn \\
&+&\frac{1}{2m_c^2}\Bigg[({\tilde \Lambda}w-{\tilde \Lambda}^\prime)\bigg(\frac{\phi_K(w)-\Delta(w){\tilde \Lambda}}{2}+\Delta_3(w)\bigg) \\
&-&\frac{1}{4}(1+w)^2\psi_2^S(w)-\frac{1}{4}(w^2-1)\psi_3^S(w) \nn \\
&+&(1+w)\left(\psi_4^S(w)-\frac{w}{2}\psi_6^S(w)-\frac{1}{4}\psi_1^A(w)-\frac{1}{2}\psi_4^A(w)\right)\nn \\
&+&\frac{2w-1}{2}\psi_5^S(w)-\frac{w-2}{2}\psi_2^A(w)+\frac{3}{4}\psi_3^A(w)\Bigg] \nn
\eea
\bea
h_T(w)&=&\Delta(w)-\left(\frac{1}{m_b}+\frac{1}{m_c} \right)\bigg(\frac{\phi_K(w)-\Delta(w){\tilde \Lambda}}{2}+\Delta_3(w) \bigg)\nn \\
&+&\frac{1}{4m_b}\bigg(3 \chi_2^A(w)+2(w-1)\chi_2^C(w) \bigg)+\frac{1}{4m_c}\bigg(3 {\bar \chi}_2^A(w)-2(w-1){\bar \chi}_2^B(w) \bigg) \nn \\
&+&\frac{1}{2m_b^2}\Bigg[({\tilde \Lambda}-w{\tilde \Lambda}^\prime)\bigg(\frac{\phi_K(w)-\Delta(w){\tilde \Lambda}}{2}+\Delta_3(w)\bigg)+\chi_1(w)\nn \\
&+&\frac{1}{4}(1+w)^2\psi_2^S(w)+\frac{1}{4}(w^2-1)\psi_3^S(w)\nn \\
&+&(1+w)\left(-\psi_4^S(w)-\frac{w}{2}\psi_6^S(w)+\frac{1}{4}\psi_1^A(w)-\frac{1}{2}\psi_4^A(w)\right)\nn \\
&+&\frac{2w-1}{2}\psi_5^S(w)+\frac{w-2}{2}\psi_2^A(w)-\frac{3}{4}\psi_3^A(w)\Bigg]\nn \\
&-&\frac{1}{2m_c^2}\Bigg[({\tilde \Lambda}w-{\tilde \Lambda}^\prime)\bigg(\frac{\phi_K(w)-\Delta(w){\tilde \Lambda}}{2}+\Delta_3(w)\bigg)-{\bar \chi}_1(w)  \\
&-&\frac{1}{4}(1+w)^2\psi_2^S(w)-\frac{1}{4}(w^2-1)\psi_3^S(w)\nn \\
&+&(1+w)\left(\psi_4^S(w)-\frac{w}{2}\psi_6^S(w)-\frac{1}{4}\psi_1^A(w)-\frac{1}{2}\psi_4^A(w)\right)\nn \\
&+&\frac{2w-1}{2}\psi_5^S(w)-\frac{w-2}{2}\psi_2^A(w)+\frac{3}{4}\psi_3^A(w)\Bigg]\nn \\
&+&\frac{1}{2m_b m_c}\Bigg[\frac{w+2}{4}\psi_1^S(w)+\frac{w^2-1}{4} \left(\psi_2^S(w)-\psi_3^S(w)+\psi_1^A(w) \right)\nn \\
&-&\frac{3}{2}(w-1) \left(\psi_4^S(w)+\psi_2^A(w) \right)-\frac{w-7}{4} \psi_3^A(w)\Bigg] \nn
\eea
\bea
h_S(w)&=&\Delta(w)-\left(\frac{1}{m_b}+\frac{1}{m_c} \right)\frac{w-1}{w+1}\bigg(\frac{\phi_K(w)-\Delta(w){\tilde \Lambda}}{2}+\Delta_3(w) \bigg)\nn \\
&+&\frac{1}{4m_b}\bigg(3 \chi_2^A(w)+2(w-1)\chi_2^C(w) \bigg)+\frac{1}{4m_c}\bigg(3 {\bar \chi}_2^A(w)-2(w-1){\bar \chi}_2^B(w) \bigg) \nn \\
&+&\frac{1}{2m_b^2}\Big[({\tilde \Lambda}-w{\tilde \Lambda}^\prime)\frac{w-1}{w+1}
\bigg(\frac{\phi_K(w)-\Delta(w){\tilde \Lambda}}{2}+\Delta_3(w) \bigg)+\chi_1(w)\Big]\nn \\
&-&\frac{1}{2m_c^2}\Big[({\tilde \Lambda}w-{\tilde \Lambda}^\prime) \frac{w-1}{w+1}
\bigg(\frac{\phi_K(w)-\Delta(w){\tilde \Lambda}}{2}+\Delta_3(w) \bigg)-{\bar \chi}_1(w)\Big]\nn \\
&+&\frac{1}{8m_b^2}(w-1)\Big[(w+1)\psi_2^S(w)+(w-1)\psi_3^S(w)-4\psi_4^S(w)+2\frac{2w-1}{w+1}\psi_5^S(w)\nn \\
&-&2w\psi_6^S(w)+\psi_1^A(w)+2\frac{w-2}{w+1}\psi_2^A(w)-\frac{3}{w+1}\psi_3^A(w)-2\psi_4^A(w) \Big] \label{hS} \\
&+&\frac{1}{8m_c^2}(w-1)\Big[(w+1)\psi_2^S(w)+(w-1)\psi_3^S(w)-4\psi_4^S(w)-2\frac{2w-1}{w+1}\psi_5^S(w)\nn \\
& +&2w\psi_6^S(w)+\psi_1^A(w)+2\frac{w-2}{w+1}\psi_2^A(w)-\frac{3}{w+1}\psi_3^A(w)+2\psi_4^A(w) \Big] \nn \\
&+&\frac{1}{8m_b m_c}\Big[(2+w)\psi_1^S(w)-(w-7)\psi_3^A(w)-6(w-1)\big[\psi_4^S(w)+\psi_2^A(w)\big]\nn \\
&+&(w^2-1)\big[\psi_2^S(w)-\psi_3^S(w)+\psi_1^A(w) \big] \Big] \nn
\eea

\subsection{$B_c \to J/\psi$}
\bea
h_V(w)&=&\Delta(w)+\frac{1}{4m_b}\left[3 \chi_2^A(w)+2(w-1)\chi_2^C(w)-4 \left( \Delta_3(w)+ \frac{\phi_K(w)-\Delta(w) {\tilde \Lambda}}{2} \right) \right] \nn \\ &-&\frac{1}{4m_c}\left[{\bar \chi}_2^A(w)+ 2 \left( \phi_K(w)-\Delta(w) {\tilde \Lambda} \right)\right] \nn \\ &+&\frac{1}{2m_b^2} \Bigg[ \left( {\tilde \Lambda} - w{\tilde \Lambda}^\prime \right) \left( \frac{\phi_K(w)-\Delta(w) {\tilde \Lambda}}{2} + \Delta_3(w) \right) + \chi_1(w)- \frac{3}{4} \psi_3^A(w) \nn \\ 
&+& (1+w) \left(\frac{1}{4} \psi_1^A(w) - \psi_4^S(w) -\frac{1}{2} \psi_4^A(w) - \frac{1}{2} w \psi_6^S(w)  \right)+\frac{w-2}{2} \psi_2^A(w) \nn \\ &+& \frac{2w-1}{2} \psi_5^S(w) +\frac{w^2-1}{4} \psi_3^S(w) + \frac{(w+1)^2}{4} \psi_2^S(w) \Bigg] \label{hpiu} \\ 
&+& \frac{1}{2m_c^2}\Bigg[ \left( {\tilde \Lambda}^\prime - w{\tilde \Lambda} \right) \frac{\phi_K(w)-\Delta(w){\tilde \Lambda}}{2} + {\bar \chi}_1(w)- \frac{1}{4} \psi_3^A(w)-\frac{1}{2} \psi_2^A(w) \nn \\ 
&+& (1+w) \left(\frac{1}{4} \psi_1^A(w) - \frac{1}{2} \psi_4^S(w) + \frac{1}{2} w \psi_6^S(w)  \right)-\frac{1}{2} w \psi_5^S(w) + \frac{w^2-1}{4} \psi_3^S(w) + \frac{(1+w)^2}{4} \psi_2^S(w) \Bigg]\nn \\ 
&+&\frac{1}{8m_b m_c}\bigg[w \psi_1^S(w)+(w^2-1)\left(\psi_2^S(w)-\psi_3^S(w)+\psi_1^A(w)\right)\nn \\ 
&+& 2 (1-w)\left(2 \psi_4^S(w)+ 2 \psi_2^A(w) + \psi_4^A + \psi_5^S \right)+(3-w)\psi_3^A(w) \bigg] \nn
\eea

\bea
h_P(w)&=&\Delta(w)+\frac{1}{4m_b}\left[3 \chi_2^A(w)+2(w-1)\chi_2^C(w)-4 \left( \Delta_3(w)+ \frac{\phi_K(w)-\Delta(w) {\tilde \Lambda}}{2}  \right) \right] \nn \\ 
&-&\frac{1}{4m_c}\left[{\bar \chi}_2^A(w)+ 4 \left( \frac{\phi_K(w)-\Delta(w) {\tilde \Lambda}}{2} - \Delta_3(w)\right)- 2(w-1) {\bar \chi}_2^B \right] \nn \\ 
&+&\frac{1}{2m_b^2} \Bigg[ \left( {\tilde \Lambda} - w{\tilde \Lambda}^\prime \right) \left(\frac{\phi_K(w)-\Delta(w){\tilde \Lambda}}{2} + \Delta_3(w) \right) + \chi_1(w)- \frac{3}{4} \psi_3^A(w) \nn \\ &+& (1+w) \left(\frac{1}{4} \psi_1^A(w) - \psi_4^S(w) -\frac{1}{2} \psi_4^A(w) - \frac{1}{2} w \psi_6^S(w)  \right)+\frac{w-2}{2} \psi_2^A(w) \nn \\ &+& \frac{2w-1}{2} \psi_5^S(w) +\frac{w^2-1}{4} \psi_3^S(w) + \frac{(w+1)^2}{4} \psi_2^S(w) \Bigg] \nn \\ 
&-& \frac{1}{2m_c^2}\Bigg[ \left( {\tilde \Lambda}^\prime - w{\tilde \Lambda} \right) \left(- \frac{\phi_K(w)-\Delta(w) {\tilde \Lambda}}{2} + \Delta_3(w) \right) \label{hP} \\
&-& {\bar \chi}_1(w) - \frac{1}{4} \psi_3^A(w) + \frac{1}{2} \psi_5^S(w) + \frac{1}{2} w \psi_2^A(w) \nn \\ &-& (1+w) \left(\frac{1}{4} \psi_1^A(w) - \frac{1}{2} \psi_4^A(w) + \frac{1}{2} w \psi_6^S(w)  \right) + \frac{1-w^2}{4} \psi_3^S(w) - \frac{(1+w)^2}{4} \psi_2^S(w) \Bigg]\nn \\ &+&\frac{1}{8m_b m_c}\bigg[(w-2)\psi_1^S(w)+(w^2-1)\left(\psi_2^S(w)-\psi_3^S(w)+\psi_1^A(w)\right)\nn \\ && + 2 (1-w)\left(2 \psi_4^A(w)+ \psi_2^A(w) + \psi_4^S(w) + 2 \psi_5^S(w) \right)-(1+w) \psi_3^A(w) \bigg] \nn
\eea
\bea
h_{A_1}&=&\Delta(w)-\frac{1}{m_b}\frac{w-1}{w+1}\bigg(\frac{\phi_K(w)-\Delta(w){\tilde \Lambda}}{2}+\Delta_3(w) \bigg)-\frac{1}{m_c}(w-1)\frac{\phi_K(w)-\Delta(w){\tilde \Lambda}}{2(w+1)} \nn \\
&+&\frac{1}{4m_b}\bigg(3 \chi_2^A(w)+2(w-1)\chi_2^C(w) \bigg)-\frac{1}{4m_c} {\bar \chi}_2^A(w)   \nn \\
&+&\frac{1}{2m_b^2} \Bigg[ \left( {\tilde \Lambda} - w{\tilde \Lambda}^\prime \right)\frac{w-1}{(w+1)} \left( \frac{\phi_K(w)-\Delta(w) {\tilde \Lambda}}{2} + \Delta_3(w) \right) + \chi_1(w)\Bigg] \nn \\
&+&\frac{1}{8m_b^2} \bigg[ (w^2-1)\psi_2^S(w)+(w-1)^2\psi_3^S(w)+2\frac{w^2-3w+2}{1+w}\psi_2^A(w) \nn \\
&+&(w-1) \big[-4\psi_4^S(w)+2\frac{2w-1}{w+1}\psi_5^S(w)-2w\psi_6^S(w)+\psi_1^A-\frac{3}{w+1}\psi_3^A(w)-2\psi_4^A(w)\big] \bigg] \nn \\
&+&\frac{1}{2m_c^2} \Bigg[ -\left( {\tilde \Lambda}w - {\tilde \Lambda}^\prime \right)\frac{w-1}{2(w+1)}  (\phi_K(w)-\Delta(w) {\tilde \Lambda})+ {\bar \chi}_1(w) \Bigg] \label{hA1} \\
&+&\frac{1}{8m_c^2} \bigg[ (w^2-1)\psi_2^S(w)+(w-1)^2\psi_3^S(w)\nn \\&+&(w-1) \big[-2\psi_4^S(w)-2\frac{w}{w+1}\psi_5^S(w)+2w\psi_6^S(w)+\psi_1^A-\frac{2}{w+1}\psi_2^A-\frac{1}{w+1}\psi_3^A(w)\big] \bigg] \nn \\
&+&\frac{1}{8m_b m_c}\bigg[w \psi_1^S(w)+(w^2-1)\big(\psi_2^S(w)-\psi_3^S(w)+\psi_1^A(w) \big)\nn \\
&+&(w-1)\big[-4\psi_4^S(w)-2\psi_5^S(w)-4\psi_2^A(w)-2\psi_4^A(w) \big]-(w-3)\psi_3^A(w) \bigg] \nn
\eea
\bea
h_{A_2}&=&\frac{1}{m_c}\bigg(\frac{1}{1+w}\big[ \phi_K(w)-\Delta(w) {\tilde \Lambda}-\Delta_3(w) \big]+\frac{1}{2}{\bar \chi}_2^B(w) \bigg) \nn \\
&+&\frac{1}{m_c^2} \frac{1}{2(w+1)}\left( {\tilde \Lambda}w - {\tilde \Lambda}^\prime \right)\big[\phi_K(w)-\Delta(w) {\tilde \Lambda}-\Delta_3(w) \big] \nn \\
&+&\frac{1}{4m_c^2} \bigg[-(1+w)\psi_2^S(w)-(w-1)\psi_3^S(w)\nn \\
&+&\psi_4^S(w)+\psi_5^S(w)-2w\psi_6^S(w)-\psi_1^A(w)+\psi_2^A(w)+\psi_4^A(w) \bigg] \label{hA2}\\
&+&\frac{1}{4m_b m_c}\bigg[-\psi_1^S(w)-(1+w)\psi_2^S(w)+(w+1)\psi_3^S(w)\nn \\
&+&3\psi_4^S(w)+3\psi_5^S(w)-(1+w)\psi_1^A(w)+3\psi_2^A(w)+\psi_3^A(w)+3\psi_4^A(w) \bigg] \nn
\eea
\bea
h_{A_3}&=&\Delta(w)-\frac{1}{m_b}\bigg( \frac{\phi_K(w)-\Delta(w){\tilde \Lambda}}{2}+\Delta_3(w)\bigg) +\frac{1}{4m_b}\bigg(3\chi_2^A(w)+2(w-1)\chi_2^C(w)\bigg)\nn \\
&-&\frac{1}{m_c(1+w)}\big[(w-1)\frac{\phi_K(w)-\Delta(w) {\tilde \Lambda}}{2}+\Delta_3(w)\big] -\frac{1}{4m_c}\big[{\bar \chi}_2^A(w)+2{\bar \chi}_2^B(w)\big]\nn \\
&+&\frac{1}{2m_b^2}\Bigg[ \left( {\tilde \Lambda} - w{\tilde \Lambda}^\prime \right)\big[ \frac{\phi_K(w)-\Delta(w) {\tilde \Lambda}}{2}+\Delta_3(w)\big] +\chi_1(w)\Bigg]\nn \\
&-&\frac{1}{2m_c^2}\Bigg[ \frac{1}{1+w}\left( {\tilde \Lambda}w - {\tilde \Lambda}^\prime \right)\big[(w-1) \frac{\phi_K(w)-\Delta(w) {\tilde \Lambda}}{2}+\Delta_3(w)\big] -{\bar \chi}_1(w)\Bigg]\nn \\
&+&\frac{1}{8m_b^2}\Big[(1+w)^2\psi_2^S(w)+(w^2-1)\psi_3^S(w)+2(2w-1)\psi_5^S(w)+2(w-2)\psi_2^A(w)\nn \\
&-&3\psi_3^A(w)+(1+w)\big(-4\psi_4^S(w)-2w\psi_6^S(w)+\psi_1^A(w)-2\psi_4^A(w)\big)\Big] \label{hA3} \\
&+&\frac{1}{8m_c^2}\Big[(w^2-1)\psi_2^S(w)+(w-1)^2\psi_3^S(w)-2w\psi_4^S(w)-2(w-1)\psi_5^S(w)\nn \\
&-&\psi_3^A(w)+2\psi_4^A(w)+(w-1)\big(2w\psi_6^S(w)+\psi_1^A(w)\big)\Big]\nn \\
&+&\frac{1}{8m_b m_c}\Big[(2+w)\big(\psi_1^S(w)-2\psi_5^S(w)-2\psi_4^A(w)\big)-2(2w+1)\big(\psi_4^S(w)+\psi_2^A(w)\big) \nn \\
&+&(1+w)^2\big(\psi_2^S(w)-\psi_3^S(w)+\psi_1^A(w)\big)-(w-1)\psi_3^A(w) \Big] \nn
\eea
\bea
h_{T_1}&=&\Delta(w)+\frac{1}{4m_b}\bigg(3 \chi_2^A(w)+2(w-1)\chi_2^C(w)\bigg)-\frac{1}{4m_c}{\bar\chi}_2^A(w)\nn \\
&+& \frac{1}{2m_b^2}\chi_1(w)+\frac{1}{2m_c^2}{\bar\chi}_1(w) \nn \\
&+&\frac{1}{8m_b m_c}\Big[-w \psi_1^S(w)-(w^2-1)[\psi_2^S(w)- \psi_3^S(w)+\psi_1^A(w)] \label{hT1} \\
&+&(w-1)[4\psi_4^S(w)+2\psi_5^S(w)+4\psi_2^A(w)+2\psi_4^A(w)]+(w-3)\psi_3^A(w) \Big] \nn
\eea
\bea
h_{T_2}&=&-\frac{1}{m_b}\Big( \frac{\phi_K(w)-\Delta(w)\tilde{\Lambda}}{2}+ \Delta_3(w)\Big)+\frac{1}{2m_c}  (\phi_K(w)-\Delta(w)\tilde{\Lambda})\nn \\
&+&\frac{1}{2m_b^2}(\tilde{\Lambda}-w\tilde{\Lambda^\prime} )\Big[\frac{\phi_K(w)-\Delta(w)\tilde{\Lambda}}{2}+ \Delta_3(w)\Big]+\frac{1}{2m_c^2}(\tilde{\Lambda} w-\tilde{\Lambda^\prime}) \frac{\phi_K(w)-\Delta(w)\tilde{\Lambda}}{2} \nn \\
&+&\frac{1}{8m_b^2}\Big[(w+1)^2 \psi_2^S(w)+(w^2-1) \psi_3^S(w)+(1+w)[-4\psi_4^S(w)-2w\psi_6^S(w)+\psi_1^A(w)-2\psi_4^A(w)]\nn \\
&+&2(2w-1)\psi_5^S(w)+2(w-2)\psi_2^A(w)-3\psi_3^A(w)\Big] \label{hT2} \\
&+&\frac{1}{8m_c^2}\Big[-(w+1)^2 \psi_2^S(w)-(w^2-1) \psi_3^S(w)+(1+w)[2\psi_4^S(w)-2w\psi_6^S(w)-\psi_1^A(w)]\nn \\
&+&2w\psi_5^S(w)+2\psi_2^A(w)+\psi_3^A(w)\Big] \nn
\eea
\bea
h_{T_3}(w)&=& -\frac{1}{2m_c}\left[ {\bar \chi}_2^B(w) + \frac{2}{1+w} \left( \Delta_3(w) - \left( \phi_K(w)-\Delta(w) {\tilde \Lambda}\right)\right)\right] \nn \\ 
&-& \frac{1}{4m_c^2}\Bigg[\frac{2}{w+1} \left( {\tilde \Lambda}^\prime - w{\tilde \Lambda} \right) \left(  \left(\phi_K(w)-\Delta(w) {\tilde \Lambda}\right) - \Delta_3(w)\right) \nn \\ 
&+& \psi_1^A(w) - \psi_2^A(w) - \psi_4^S(w) - \psi_4^A(w) \nn \\ &+& ( w - 1 ) \psi_3^S(w) - \psi_5^S(w) + ( 1 + w ) \psi_2^S(w) + 2 w \psi_6^S(w) \Bigg]  \\ 
&-&\frac{1}{4m_b m_c}\bigg[ \psi_1^S(w) - \psi_3^A(w) -  3 \left(\psi_2^A(w) + \psi_4^A(w) + \psi_4^S(w) + \psi_5^S(w) \right) \nn \\ 
&+& (1+w)\left(\psi_1^A(w) + \psi_2^S(w) - \psi_3^S(w)\right)\bigg] \,\, . \nn
\eea
\section{Fit of the form factors $h_{T_i}$ and $h_P$ } \label{appB}
The form factors $h_{T_i}$ and $h_P$ depicted in Fig.~\ref{plotPsiFF} are fitted using the  parametrization  adopted by the HPQCD Collaboration   \cite{Cooper:2020wnj}.
The variable $z(q^2,t_0)$ is defined:
\be
z(q^2,t_0)=\frac{\sqrt{t_+-q^2}-\sqrt{t_+-t_0}}{\sqrt{t_+-q^2}+\sqrt{t_+-t_0}}
\ee
with  $t_+=(m_B+m_{D^*})^2$ and $t_0=t_-=(m_{B_c}-m_{J/\psi})^2$. 
The  form factors $h_i(q^2)$ are fitted using a truncated expansion in powers of  $z$:
\be
h_i(q^2)=\frac{1}{{P}(q^2)}\sum_{n=0}^N a_n z^n \label{FFpar}
\ee
where the Blaschke factors $P(t)$ account for the  resonances  in the $t$-channel:
\be
{P} (q^2)=\prod_{P_i} z(q^2,M_{P_i}^2) \label{eq:blaschke}.
\ee
 $h_{T_3}$ involves  poles with $J^P=1^+$,  $h_{T_{1,2}}$ with $J^P=1^-$,  $h_P$  with $J^P=0^-$, with   masses in Table \ref{tab:MP}.
\begin{table}[t]
\caption{ \small  Mass $M_{P_i}$ (in GeV) of the $\bar b c$ resonances $P_i$ included in the Blaschke factors \eqref{eq:blaschke}.}\label{tab:MP}
\vspace{0.3cm}
\centering
\begin{tabular}{c c c c }
\hline
$J^P/i$&$1^-$ &  $1^+$ & $0^-$  
\\
\hline
1&6.336 & 6.745 & 6.275 \\
2&6.926 & 6.75 & 6.872 \\
3&7.02 & 7.15 & 7.25 \\
4&7.28 & 7.15 & \\
\hline
\end{tabular}
\end{table}
%
The values of $a_n$ resulting from  the fit are collected in Table \ref{tab:an}.
\begin{table}[h]
\caption{ \small  Parameters $a_n$  in Eq.~(\ref{FFpar})}\label{tab:an}
\vspace{0.3cm}
\centering
\begin{tabular}{l |cccc }
\hline
&$a_0$ &  $a_1$ & $a_2$  & $a_3$ 
\\
\hline
$h_{T_1}$ & $0.058 \pm 0.002$ & $-0.476 \pm 0.001$ & $1.228\pm 0.131$ & $3.419 \pm 3.977$ \\
$h_{T_2}$ &$-0.021 \pm 0.003$& $0.096 \pm 0.014$ &$0.030 \pm 0.665 $ & $ 0.473 \pm 0.670$ 
 \\
$h_{T_3}$ &$-0.047 \pm 0.014$ &$0.675 \pm 0.165$ & $-0.701 \pm 1.693$ &  $0.894\pm 7.468$  \\
$h_P$ &$0.142  \pm 0.008$ & $-0.944 \pm 0.053$ & $0.157 \pm 1.629$  &$-1.495\pm 1.131$ \\
\hline
\end{tabular}
\end{table}

\bibliographystyle{JHEP}
\bibliography{refFFMNP}
\end{document}